\documentclass{aa} 
\usepackage{natbib}
\bibpunct{(}{)}{;}{a}{}{,} 
\usepackage{graphicx}
\usepackage{txfonts}
\usepackage{lscape}
\usepackage{color}
\usepackage{hyperref}
\hypersetup{
    colorlinks=blue,
    linkcolor=blue,
    filecolor=magenta,      
    urlcolor=cyan,
}
 
\usepackage{float}

\begin{document}

   \title{Herschel observations of the circumstellar environment of \\
   the two Herbig Be stars R Mon and PDS27
   \thanks{{\it Herschel} is an ESA space observatory with science instruments provided by European-led Principal Investigator consortia and with important participation from NASA.}
   }

   \author{M. J. Jim\'enez-Donaire\inst{1,2}
          \and
          G. Meeus\inst{2,3}
          \and
          A. Karska\inst{4}
          \and
          B. Montesinos\inst{5,3}
          \and 
          J. Bouwman\inst{6}
          \and 
          C. Eiroa\inst{2,3}
          \and 
          Th. Henning\inst{6}
          }

   \institute{Zentrum fur Astronomie der Universitat Heidelberg, Inst. f\"ur Theor. Astrophysik, Albert-Ueberle Str. 2, 69120 Heidelberg, Germany.
              \email{m.jimenez@zah.uni-heidelberg.de}
        \and
             Universidad Aut\'onoma de Madrid, Dpto. F\'isica Te\'orica, Facultad de Ciencias, Campus Cantoblanco, 28049 Madrid, Spain.
        \and 
            Unidad Asociada CAB--UAM, Madrid, Spain.
        \and
             Centre for Astronomy, Nicolaus Copernicus University, Faculty of Physics, Astronomy and Informatics, Grudziadzka 5, PL-87100 Torun, Poland.
        \and
             Departamento de Astrof\'{\i}sica, Centro de Astrobiolog\'{\i}a
            (CAB, CSIC-INTA), ESAC Campus, Camino Bajo del Castillo s/n,
            E-28692 Villanueva de la Ca{\~n}ada, Madrid, Spain.
        \and 
             Max Planck Institut für Astronomie, Konigstühl 17, 69117 Heidelberg, Germany.
        }


 
\abstract
{The circumstellar environment of Herbig Be stars in the far-infrared is poorly characterised, mainly because they are often embedded and rather distant. The analysis of far-infrared spectroscopy allows us to make a major step forward by covering multiple rotational lines of molecules, e.g. CO, that are useful probes of physical conditions of the gas.}
{To characterise the gas and dust in the disc and environment of Herbig Be stars, and to compare the results with those of their lower-mass counterparts, the Herbig Ae stars.}
{We report and analyse far-infrared observations of two Herbig Be stars, R Mon and PDS 27, obtained with Herschel's instruments PACS and SPIRE. 

We construct spectral energy distributions and derive the infrared excess. We extract line fluxes from the PACS and SPIRE spectra and construct rotational diagrams in order to estimate the excitation temperature of the gas. We derive CO, [O\,I] and [C\,I] luminosities to determine physical conditions of the gas, as well as the dominant cooling mechanism.}
{We confirm that the Herbig Be stars are surrounded by remnants from their parental clouds, with an IR excess that mainly originates in a disc. In R Mon we detect [O\,I], [C\,I], [C\,II], CO (26 transitions), water and OH, while in PDS 27 we only detect [C\,I] and CO (8 transitions). We attribute the absence of OH and water in PDS 27 to UV photo-dissociation and photo-evaporation. From the rotational diagrams, we find several components for CO: we derive $T_{\rm{rot}}$ 949$\pm$90 K, 358$\pm$20 K \& 77$\pm$12 K for R Mon, 96$\pm$12 K \& 31$\pm$4 K for PDS 27 and 25$\pm$8 K \& 27$\pm$6 K for their respective compact neighbours. The forsterite feature at 69 $\mu$m was not detected in either of the sources, probably due to the lack of (warm) crystalline dust in a flat disc. We find that cooling by molecules is dominant in the Herbig Be stars, while this is not the case in Herbig Ae stars where cooling by [O\,I] dominates. Moreover, we show that in the Herbig Be star R Mon, outflow shocks are the dominant gas heating mechanism, while in Herbig Ae stars this is stellar.}
{The outflow of R Mon contributes to the observed line emission by heating the gas, both in the central spaxel/beam covering the disc and the immediate surroundings, as well as in those spaxels/beams covering the parabolic shell around it. PDS 27, a B2 star, has dispersed a large part of its gas content and/or destroyed molecules; this is likely given its intense UV field. }


\titlerunning{Herschel observations of R Mon and PDS 27}
\authorrunning{M. J. Jim\'enez-Donaire et al.}

\maketitle

\section{Introduction}

Low- and intermediate-mass star formation theory assumes that 
stellar masses are accumulated as a consequence of gravitational collapse of a dense 
molecular core, while conservation of angular momentum forces the infalling material to 
form an accretion disc around the central protostar. However, once the disc is formed, 
turbulence and viscosity contribute to the motion of the disc material towards the star while 
angular momentum causes the opposite movement. During the accretion phase, viscous dissipation 
also results into the heating of the inner part of the disc, but once accretion comes to a halt, 
the disc is heated by stellar irradiation, being warm at the surface and colder in the midplane \citep{mckee2007ARA&A..45..565M}. This passive disc phase lasts for several Myrs during which 
the disc dissipates slowly due to low-level accretion, photo-evaporation and/or planet formation.

Herbig Ae/Be (HAeBe) stars are pre-main sequence A- or B-type stars with optical emission lines and an IR excess due to the thermal emission of dust grains and photon emission by polycyclic aromatic hydrocarbons (PAHs). In particular, Herbig Be (HBe) stars are the link between massive protostars and the intermediate-mass Herbig Ae (HAe) stars.

HAe discs have been characterised in terms of their $\lambda F_{\lambda}$ spectral energy distribution (SED): group I sources have a relatively strong far-IR flux, which is energetically
similar to the flux in the near-IR, while group II sources show a similar near-IR excess as
group I sources but their flux falls off more strongly towards the far-IR \citet{meeus2001A&A...365..476M}. These differences were interpreted as a consequence of a different
disc geometry, with flaring/flat discs in which dust grains grow and settle towards the
midplane \citep{dullemond2004A&A...421.1075D} and/or flat discs that are shadowed by a puffed up inner
wall \citep{dullemond2004A&A...417..159D}. However, recent observations suggest that group I sources have a dust-depleted gap in their inner disc \citep{maaskant2013A&A...555A..64M} beyond which the disc flares.

For the more massive B and O type stars, however, the picture of a disc is less clear. The 
evolution of their discs has been theoretically studied \citep{hollenback1994ApJ...428..654H, gorti2009ApJ...690.1539G}, and it is predicted that their strong UV radiation evaporates part 
of it on a timescale $\sim 10^{5}$ yrs, so that they are smaller and less massive than HAe discs. Based on millimetre observations, \citet{fuente2003ApJ...598L..39F} concluded that the discs around HBe stars are likely flat and loose a large fraction of their mass (90\%) before the pre-main-sequence stellar phase. Also IR interferometric data suggest that the discs of HBe stars are flatter and truncated on 
the outside, as seen for MWC 147 \citep{kraus2008A&A...489.1157K} and MWC 297
\citep{acke2008A&A...485..209A}. The flattening of the disc could be due to rapid grain growth, causing the optical depth of the disc to drop and allowing the UV photons to penetrate deeper 
into the disc and photo-evaporate its external layers \citep{dullemond2004A&A...421.1075D}, while dust settling occurs at a higher rate.

\citet{verhoeff2012A&A...538A.101V} selected a sample of HBe stars to study the spatial distribution and mineralogy of their warm dust, and compared the results to HAe stars. Their mid-IR imaging revealed compact material distributed in discs, and more diversity in the shape of the infrared SEDs of HBe stars when compared to HAe stars. They also found that most HBe discs are more flat and truncated on the outside, so not just scaled-up versions of HAe discs. 

In this paper, we will analyse two HBe objects, R Mon and PDS 27, for which we have {\em Herschel} PACS \citep[Photodetector Array Camera and Spectrometer; ][]{poglitsch2010A&A...518L...2P} and SPIRE \citep[Spectral and Photometric Imaging REceiver;][]{griffin2010A&A...518L...3G} far-IR spectra, as well as photometric images. In Section~\ref{s_sample} we list the properties of our stars, while in Section~\ref{s_obs} we present the observations. The results and discussion are shown in Sections~\ref{s_results} and ~\ref{s_discussion}, while we round off with conclusions in  and~\ref{s_conclusions}.

\section{Sample}
\label{s_sample}

\begin{table*}
\caption{Stellar parameters of the HBe stars.}
\label{t_para}
\centering
\begin{tabular}{lccccccccc}
\hline\hline
Name & RA & DEC & Spec. & log $T_{eff}$ & Distance & $A_{V}$ & log $L_{*}$ & Radius & Mass\\
 &  \multicolumn{2}{c}{J2000.0}  & Type & (K) & (kpc) & (mag) & ($L_{\odot}$) & ($R_{\odot}$) & ($M_{\odot}$)\\
\hline
R Mon & 06h 39' 09.9\arcsec & +08$^{\circ}$ 44' 10.7\arcsec & B8IIIev & 4.07 $\pm$ 0.05 & 0.8 & 2.5 $\pm$ 0.1 & 2.12 $\pm$ 0.12 & 2.8 $\pm$ 0.4 & 3.4 $\pm$ 0.8\tablefootmark{a}\\

PDS 27 & 07h 19' 35.9\arcsec & -17$^{\circ}$ 39' 18.7\arcsec & B2?e\tablefootmark{b} & 4.34 $\pm$ 0.14 & 1.25\tablefootmark{d} & 4.8 $\pm$ 0.1 & 3.80 $\pm$ 0.11 & 5.5 $\pm$ 0.8 & 9.1 $\pm$ 1.8\tablefootmark{c}\\
\hline
\end{tabular}
\tablefoot{
\tablefoottext{a}{\citet{mora2001A&A...378..116M}, }
\tablefoottext{b}{\citet{vieira2003AJ....126.2971V}, and }
\tablefoottext{c}{\citet{jones1982AJ.....87.1223J}}.
\tablefoottext{d}{\citet{vieira2003AJ....126.2971V}; \citet{ababakr2015MNRAS.452.2566A} 
give a distance of 3.17 kpc, based on kinematic arguments.}
}
\end{table*}

R Mon and PDS 27 are part of a larger sample of Herbig Be stars in a {\em Herschel} proposal (OT1\_gmeeus\_1). However, these are the only two objects that were observed in spectroscopic mode, allowing us to study the gas and dust properties. We list their parameters in Table~\ref{t_para}.

\paragraph{R Mon} is an intermediate-mass pre-main-sequence star with a spectral type B8IIIev \citep{mora2001A&A...378..116M}, classified as a HAeBe star \citep{herbig1960ApJS....4..337H, finkenzeller1984A&AS...55..109F}. It is located at a distance of 800 pc \citep{jones1982AJ.....87.1223J} and its apparent V magnitude varies between 11 and 13.8 mag. CO radio observations ($^{12}$CO $J=1\rightarrow0$) showed that R Mon is located in a flattened
molecular cloud and drives a strong bipolar outflow in north-south direction, that has carved out a bipolar cavity \citep{canto1981ApJ...244..102C} where the southern part is being obscured. The walls of the cavity are illuminated by R Mon, creating "Hubble's Variable Nebula" or NGC 2261, extending 2' to the North along P.A. 350$^{\circ}$ \citep{close1997ApJ...489..210C}.

\citet{brugel1984ApJ...287L..73B} detected a high-velocity ($\sim$100 km s$^{-1}$) collimated 
jet in [SII], positioned through the axis of symmetry of the reflection nebula NGC 2261. The jet extends 4-10$\arcsec$ \ to the North and 10-16$\arcsec$ \ to the South. They concluded that the collimation must occur close ($\sim 2-3\arcsec$, or 2000 au) to the star. High-resolution (FWHM
0.2$\arcsec$) adaptive optics observations by \citet{close1997ApJ...489..210C} uncovered a 
companion at a separation of 0.69$\arcsec$ \ from R Mon, thought to be a very young T Tauri star
(TTS). Scattered light images show a conical reflection nebula. 
\citet[see Figure 15]{close1997ApJ...489..210C} proposed a model in which R Mon has a small 
($< 100$ au) disc surrounded by an optically thick, dusty flattened envelope. A strong outflow 
carves out a conical cavity in the surrounding envelope, that is optically thin. 

\citet{fuente2003ApJ...598L..39F} derived from 2.7mm interferometric observations that the 
emission is extended up to 3-4$\arcsec$, thus $\sim$ 3000 au, too large for a disc, and also
suggested a flattened envelope surrounding the disc. A few years later, \citet{fuente2006ApJ...649L.119F} presented PdBI continuum observations at 1.3 and 2.7mm with 
higher angular resolution and sensitivity, as well as $^{\rm{12}}$CO rotational lines. The 1.3mm
image shows a disc extending up to 0.3$\arcsec$, with a mass of 0.007 $M_{\odot}$. The CO\,(2-1) and CO\,(1-0) lines are optically thick, and consistent with a gaseous 
disc in keplerian rotation with an outer radius of 1500 au and a total disc mass of 0.014--0.08 $M_{\odot}$. However, \citet{sandell2011ApJ...727...26S} argue that their CO rotation curve might 
be contaminated by emission from the outflow. \citet{alonso-albi2009A&A...497..117A} also suggests 
that a gaseous flat disc surrounds R Mon, based on their $^{\rm{13}}$CO observations. 

\paragraph{PDS 27.} Also known as DW CMa, is located on the edge of the dark cloud KHAV 201. Its 
spectral type is estimated to be B2?e \citep{vieira2003AJ....126.2971V}.
\citet{manchado1990A&AS...84..517M} observed P-Cygni profiles and concluded that they could be reproduced with a wind velocity of about 130 $\pm$ 30 km s$^{-1}$, typical for optical jets,  
Herbig-Haro (HH) and solar-mass TTS. 

The presence of hot dust with a temperature of up to 1400 K can be derived from its location 
in the near-IR colour-colour diagram (J--H vs. H--K), whereas far-IR data also indicate the 
presence of cold dust with temperatures ranging from 60 to 200 K 
\citep{beichman1986ApJ...307..337B, manchado1990A&AS...84..517M}. 
There are no millimetre observations for PDS 27 in the literature, nor a mention of a close companion.

\section{Observations and data reduction}
\label{s_obs}

We obtained PACS and SPIRE spectroscopic data for R Mon and PDS 27. The data were reduced with the official version 14.0 of the Herschel Interactive Processing Environment \citep[HIPE;][]{ott2010ASPC..434..139O}. 

PACS is an integral field spectrometer (IFS) with a 5$\times$5 array of spectral pixels, each 9.4\arcsec$\times$ 9.4\arcsec
in size (also referred to as \textquotedblleft{}spaxels\textquotedblright{}). They cover the spectral range from 51 to 210 $\mu$m 
with a resolution $\lambda/\Delta\lambda \sim \, 1000-3000$, divided into four orders, referred to as 
\textquotedblleft{}B2A\textquotedblright{}, \textquotedblleft{}B2B\textquotedblright{},
\textquotedblleft{}short R1\textquotedblright{} and \textquotedblleft{}long
R1\textquotedblright{}. For each spectroscopic observation, two settings were used in order to have a full
coverage: SED B2A + short R1 (51-73 $\mu$m + 102-146 $\mu$m) and SED B2B + long R1 (70-105 $\mu$m + 
140-220 $\mu$m). Two repetitions were obtained in order to be able to remove cosmic 
rays and improve the wavelength accuracy. The total observation time adds up to 7037 seconds for each target. 
Each order was reduced using a modified pipeline optimised for extended sources \citep{green2013ApJ...770..123G}. We extracted the spectra from the central spaxel, scaled to the flux of the central 3$\times$3 spaxels (see Figure \ref{f_CII}). 

SPIRE spectroscopic data were taken in a single pointing with sparse spatial sampling, at high 
(H) spectral resolution. For each source, 20 repetitions were performed, adding up to a total
observation time of 3423 seconds per target. The spectrum is divided into two orders covering 
the spectral ranges 194\textendash{}325 $\mu$m (\textquotedblleft{}SSW\textquotedblright{},
Spectrograph Short Wavelengths) and 320\textendash{}690 $\mu$m
(\textquotedblleft{}SLW\textquotedblright{}, Spectrograph Long Wavelengths), with a resolution 
of $\lambda/\Delta\lambda \sim \, 300-800$, increasing at shorter wavelengths. Each order was
reduced separately within HIPE, using the standard pipeline for single pointing. We first extracted 
the spectra from the SPIRE detectors SSW D4 and SLW C3, which are centered on the main sources, 
R Mon and PDS 27. We also extracted the other detectors to analyse the surrounding environment. 

The fluxes were extracted using a gaussian fit to the 
lines with a first-order polynomial to the continuum. We used the RMS on the continuum 
(excluding the line) to derive a 1$\sigma$ error on the line by integrating a gaussian with 
height equal to the continuum RMS and width of the instrumental FWHM. We only call a line a
detection when it has at least a 3 $\sigma$ signal. The SPIRE spectra for R Mon and PDS 27 in Figure \ref{f_specSPIRE} show, especially for R Mon, absorption-like features near the detected emission lines. These features are due to the instrumental line shape of the FTS instrument, which is an asymmetric {\it sinc} function.

\begin{figure}
\includegraphics[scale=0.4]{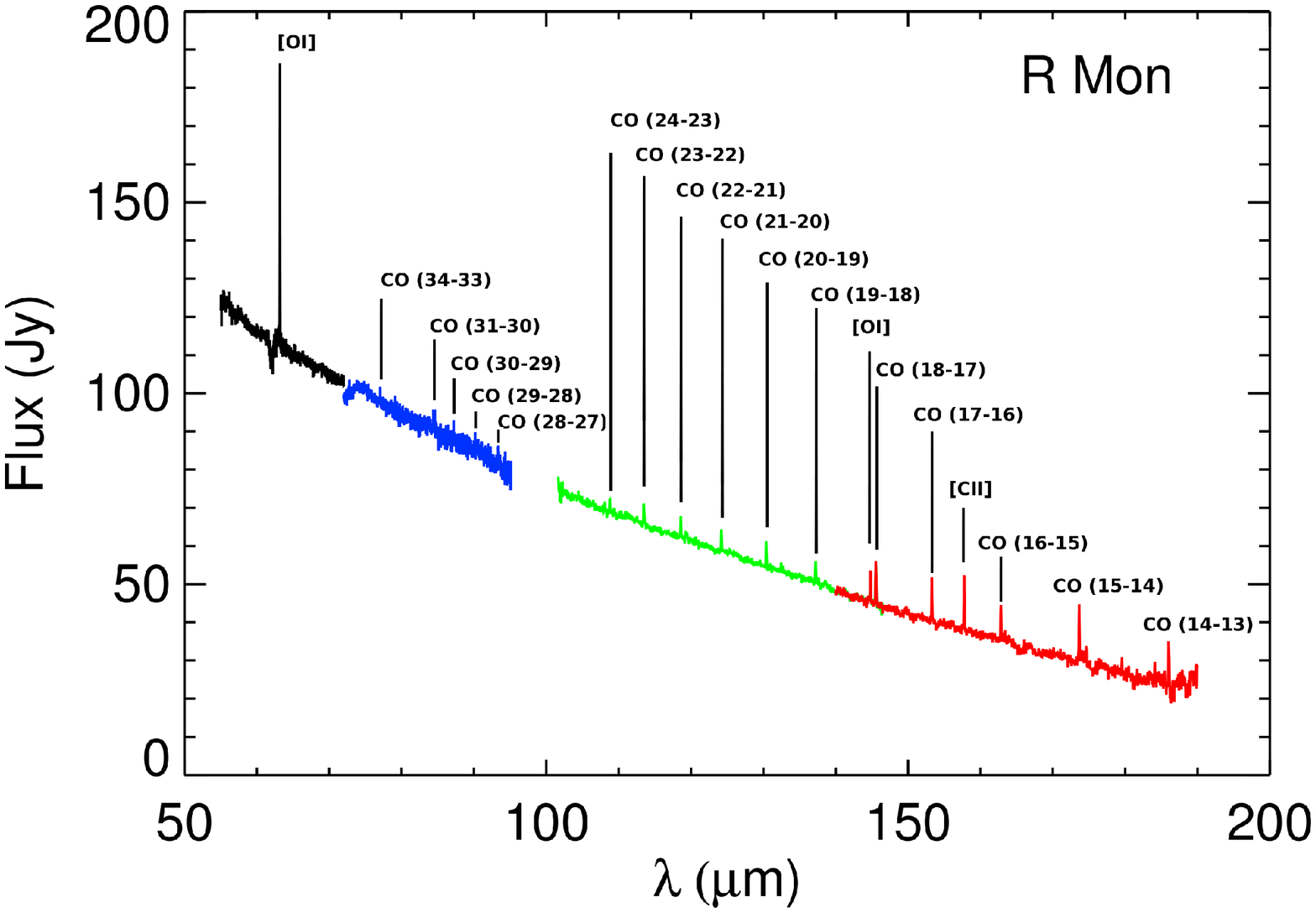}
\includegraphics[scale=0.4]{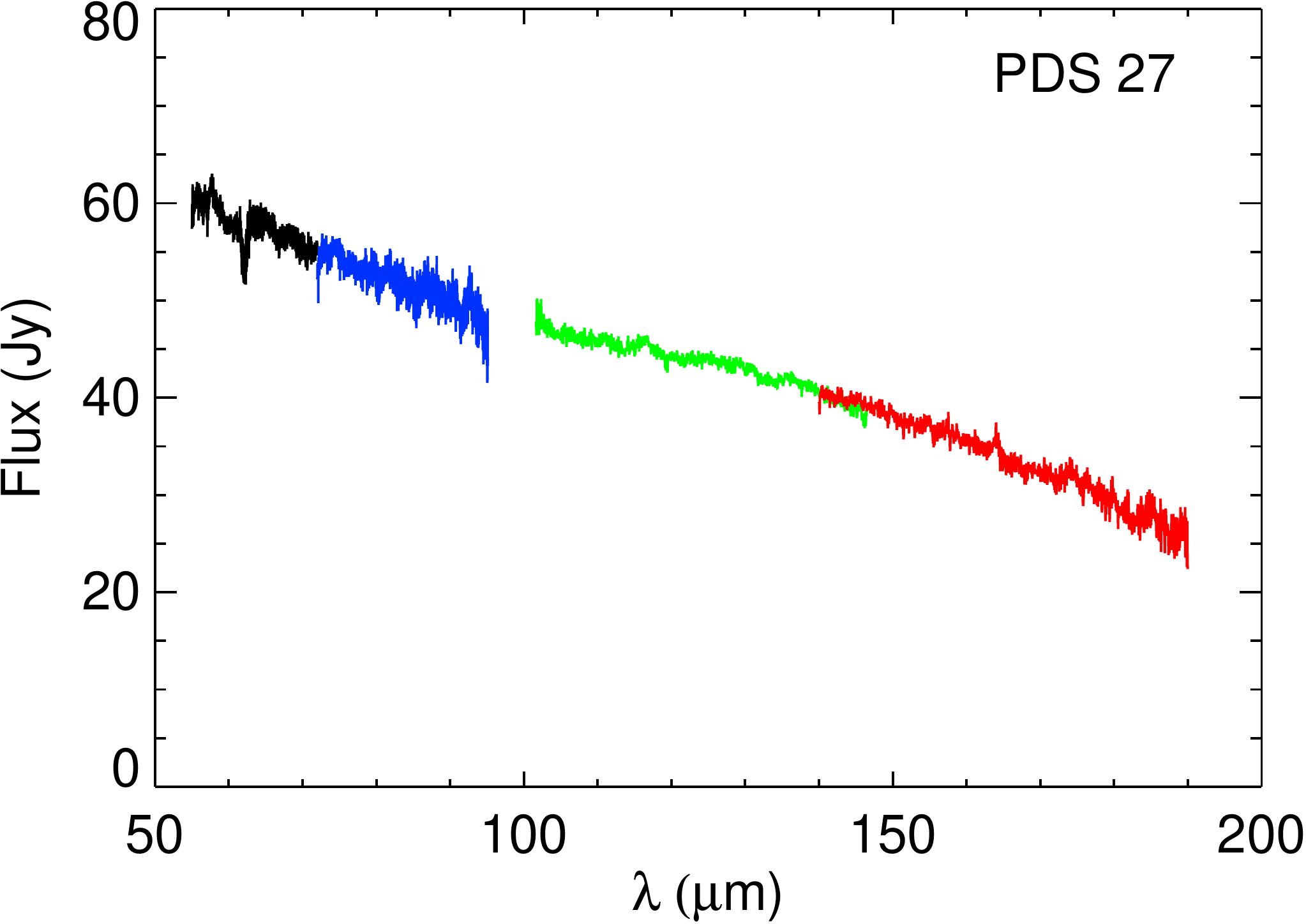}
\caption{PACS spectra of R Mon (top) and PDS 27 (bottom). The different segments have 
been scaled to match the shortest wavelength.}
\label{f_specPACS}
\end{figure}

\begin{figure}
\includegraphics[scale=0.4]{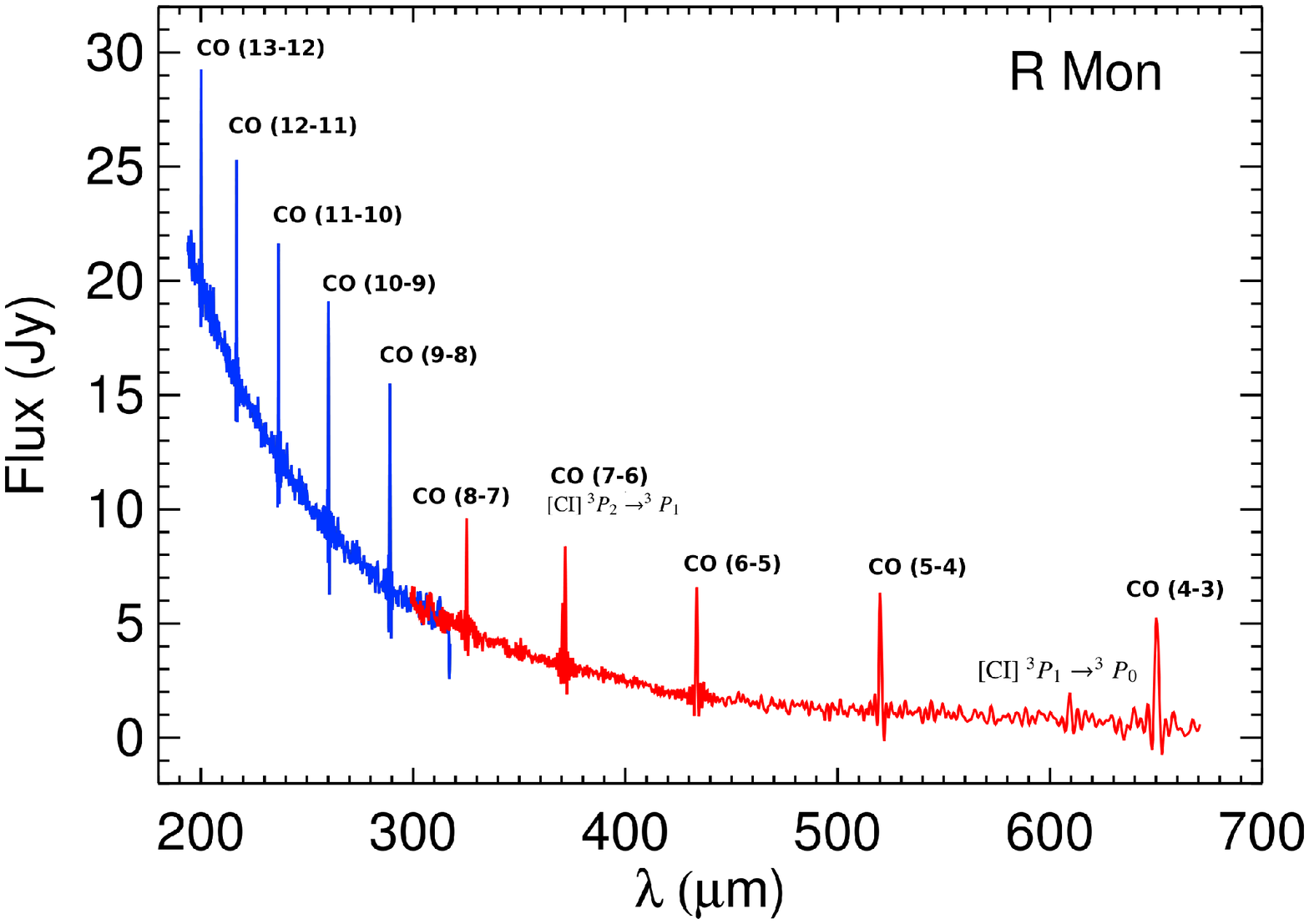}
\includegraphics[scale=0.4]{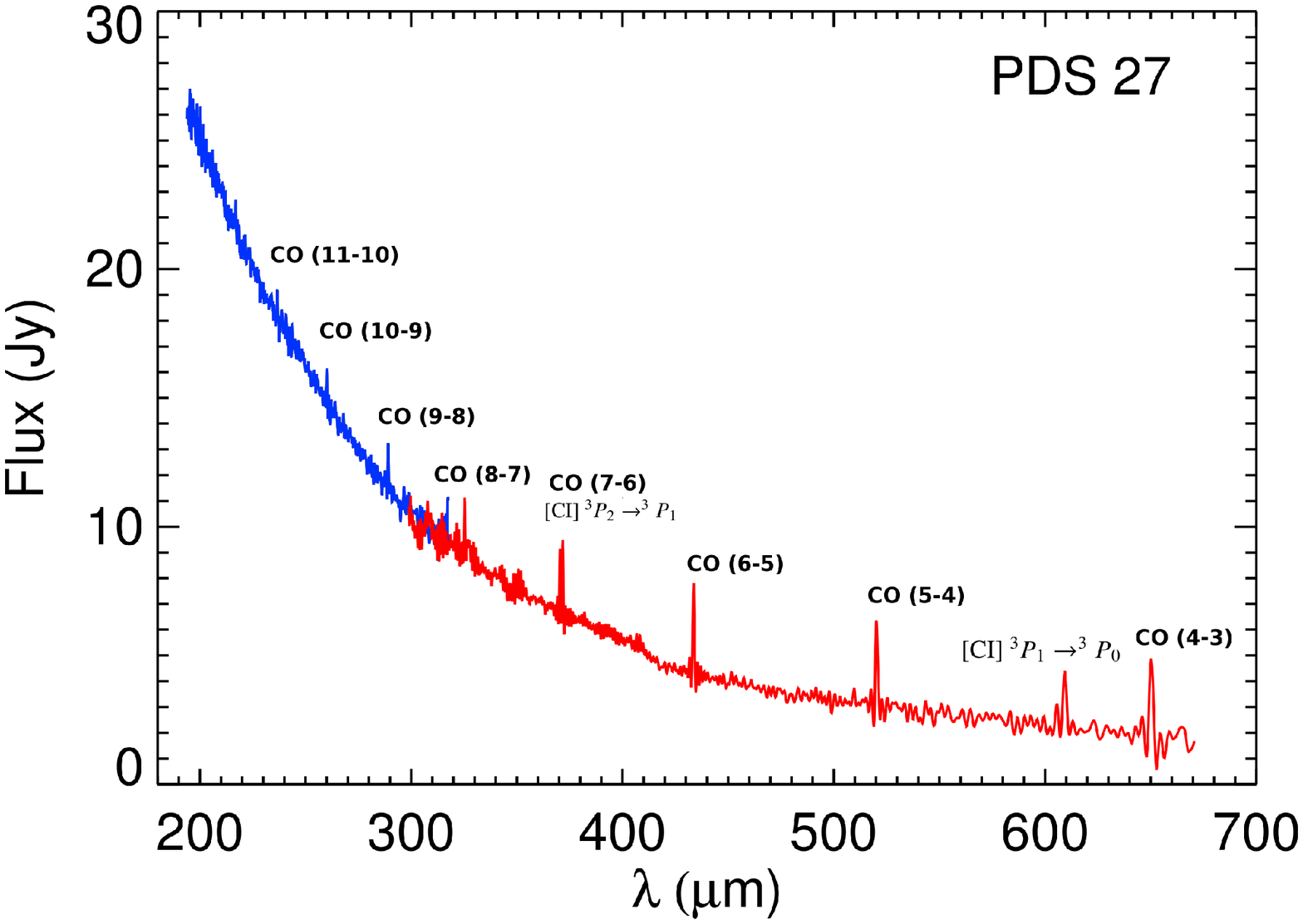}
\caption{SPIRE spectra of R Mon (top) and PDS 27 (bottom). The different segments have 
been scaled to match the PACS spectrum.} 
\label{f_specSPIRE}
\end{figure}

Without any prior knowledge about the size
of the objects, we derived that our sources are partially extended in the SPIRE beams by comparing 
the point-source and extended-source calibration available in the pipeline. Therefore, we applied a correction using the 'semiExtendedCorrector' task in HIPE, in order to extract the spectra in the
central beam. The output is a spectrum corrected to a reference beam size that is constant with frequency (see Section 7.6 of the SPIRE Data Reduction Guide - DRG).

Both sources are plagued with surrounding emission that contributes more towards longer wavelengths where larger beams are used. Given the difference in spaxel/beam size between PACS and SPIRE, we can expect an offset in flux levels between the spectra at the wavelengths where they overlap. Therefore, we first produced a continuous spectrum over the whole 50-670 $\mu$ wavelength. By doing this we assume that the emission comes from a region similar in size over the whole wavelength range. This method is commonly used for non-compact sources \citep[e.g.,][]{green2013ApJ...770..123G}. In order to produce a continuous spectrum from 50 to 670 micron, we first scaled the PACS bands so they would match, taking the shortest wavelength range (most compact) as a reference. Next we matched the SPIRE bands with those of PACS. The factors used to match the range 50-75 $\mu$m are 1, 0.93, 0.91, 0.87, 0.9 for R Mon and 1.04, 0.95, 0.94, 0.77, 0.75 for PDS 27 (ranges 70-95, 100-150, 140-200, 200-300, 280-670 $\mu$m, respectively). The resulting scaled spectra are shown in Figure~\ref{f_specPACS} and~\ref{f_specSPIRE}, and used to measure the line fluxes.

\section{Results}
\label{s_results}

\subsection{Continuum SEDs}

\begin{figure}
\includegraphics[scale=0.4]{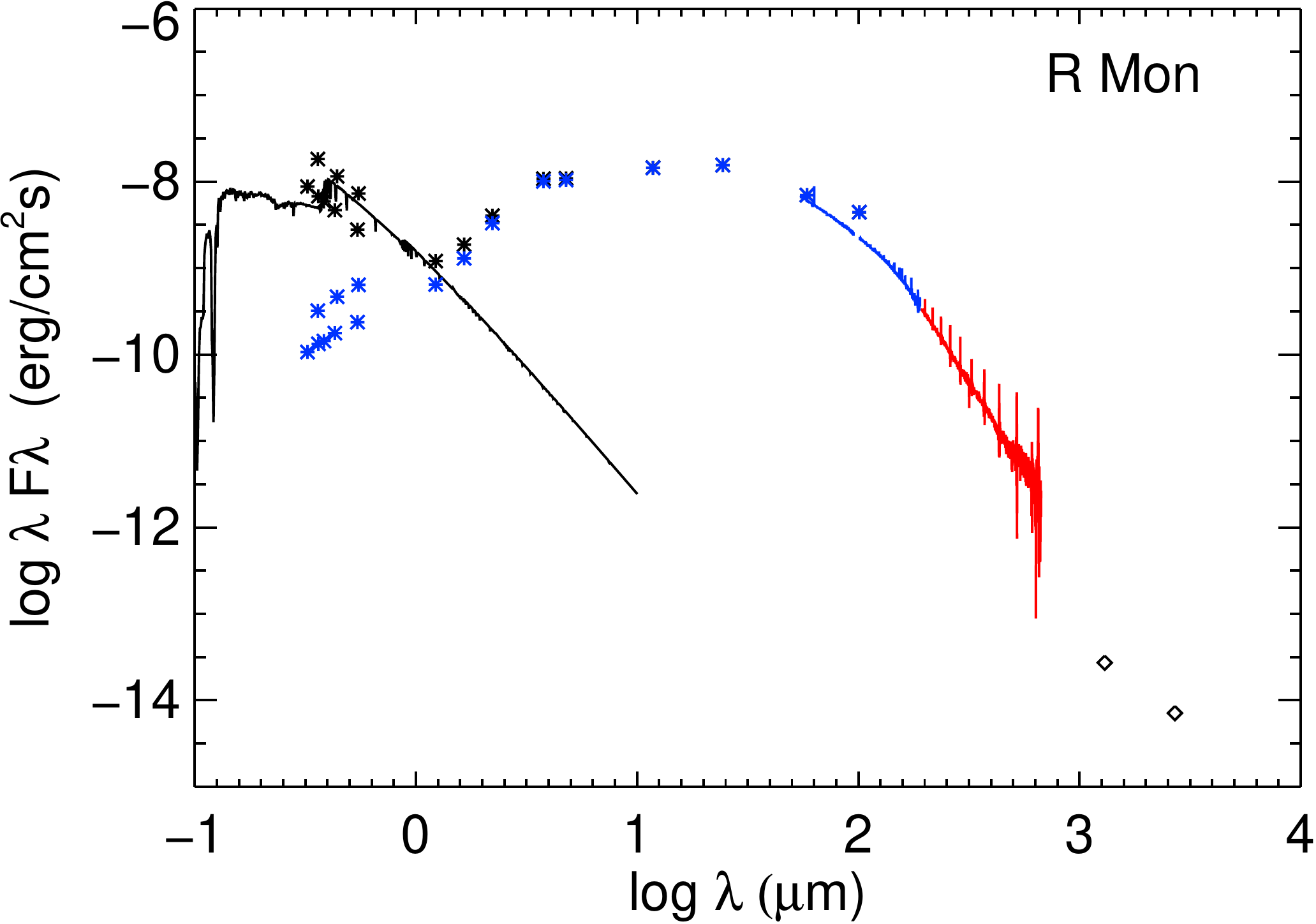}
\includegraphics[scale=0.4]{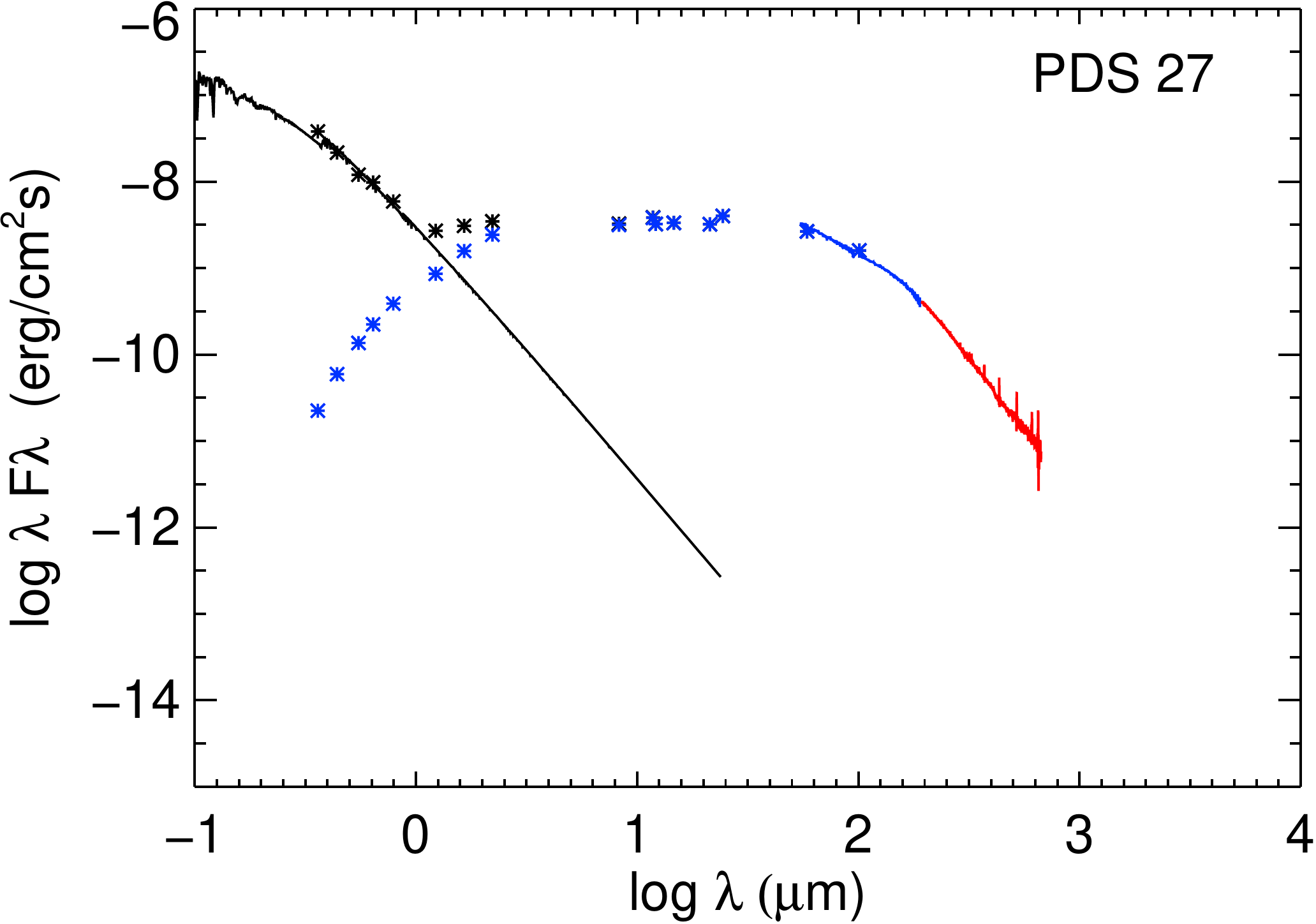}
\caption{SED for R Mon (top) and PDS 27 (bottom), including our new PACS and SPIRE spectroscopy (red curves). We show both the dereddened (black points) and observed (blue points) fluxes. Overplotted are the Kurucz models (black curves). For R Mon we show the fluxes both in its brighter and fainter state. The millimetre data from \citet{fuente2003ApJ...598L..39F} is shown in the upper panel as diamonds.}
\label{f_SED}
\end{figure}

The ancillary optical and near-IR photometric data \citep{verhoeff2012A&A...538A.101V} was 
complemented with our new Herschel spectroscopic data, and the millimeter data from \mbox{\citet{fuente2003ApJ...598L..39F}} for R Mon, in order to construct the spectral energy distribution (see Figure~\ref{f_SED}).

As seen in Figure~\ref{f_SED}, our sources have a disc-like IR excess that starts in the near-IR and extends to the far-IR/mm region, with a maximum around 30 $\mu$m. This excess can in part be attributed to the emission from the disc. In order to compare the IR excess to the stellar emission, we 
calculated the ratio between the observed emission and the Kurucz atmosphere model \citep{kurucz1993yCat.6039....0K}, both for the whole wavelength range and in the near-, mid- and far-IR (see Table~\ref{t_excess}). The total IR excess L$_{\rm{IR}}$/L$_{*}$ is $\sim$ 3.36 for R Mon and $\sim$ 0.13 for PDS 27, rather different from HAe discs that have on average 0.57 and 0.31 for group I and II, respectively \citep{pascual2016A&A...586A...6P}. This could be attributed to differences in their discs properties. A possible explanation for the large excess ratio seen in R Mon could be the presence of an outflow; systems which exhibit strong outflows have discs that are powered by accretion \citep{2012ApJ...758..100B}. It is very likely that PDS 27 has a small depleted disc, reprocessing the stellar radiation.

The FIR-to-NIR ratio is useful to provide an idea of the morphology of the protostellar disc. The ratios we compute are 2.17 for R Mon and 1.73 for PDS 27, comparable to average values of HAe discs, 2.27 for group I (flared) and 1.17 for group II (flat) discs \citep{pascual2016A&A...586A...6P}. We conclude that the HBe values are indicative of flared discs.

\begin{table}[]
    \caption{IR excesses and excess ratio calculated for both stars. }
    \label{t_excess}
\begin{tabular}{lccccc}
\hline
\hline
  & L$_{\rm{IR}}$/L$_*$ &L$_{\rm{NIR}}$/L$_*$ &L$_{\rm{MIR}}$/L$_*$ & L$_{\rm{FIR}}$/L$_*$ & L$_{\rm{FIR}}$/L$_{\rm{NIR}}$\\
        & (total) & 2--5     & 5--20     & 20--200  & \\
        &         & ($\mu$m) & ($\mu$m)  & ($\mu$m) & \\
\hline
R Mon           & 3.36   & 0.57  &1.52   & 1.21   &2.17\\
PDS 27          & 0.13   & 0.03  &0.04   & 0.05   &1.73\\
group I              & 0.57   & 0.13  &0.11   & 0.21   &2.27\\
group II             & 0.31   & 0.10  &0.09   & 0.05   &1.17\\
\hline
    \end{tabular}
{Notes: HAe group I and II average values taken from \citet{pascual2016A&A...586A...6P}.}
\end{table}

\subsection{Line emission}

\begin{table}
\caption{Line fluxes of the detected emission lines, and their 1 $\sigma$ error. $^a$ blend
of OH and CO. Lines at $\lambda \ >$ 200 $\mu$m were observed with SPIRE, thus 
have a larger beam than those observed with PACS ($\lambda \ <$ 200 $\mu$m). }
\label{t_emi_lines}      
\begin{tabular}{llcc}       
\hline\hline                 
Transition & $\lambda$& \multicolumn{2}{c}{Line flux ($10^{-17}$ W m$^{-2}$)}\\
           &($\mu$m)   & R Mon & PDS 27 \\
\hline                       
[O\,I] $^{3}P_{1}\rightarrow^{3}P_{2}$ &  63.2 & 242.1$\pm$0.5 & $<$ 19.9\tabularnewline

[O\,I] $^{3}P_{0}\rightarrow^{3}P_{1}$ & 145.5 & 24.4$\pm$0.5  & $<$ 5.1\tabularnewline

[C\,II] $^{2}P_{3/2}\rightarrow^{2}P_{1/2}$ & 157.7 & 24.4$\pm$0.5 & $<$ 5.2\tabularnewline

[C\,I] $^{3}P_{2}\rightarrow^{3}P_{1}$ & 370.8 & 3.2$\pm$0.3   & 3.1$\pm$0.4\tabularnewline

[C\,I] $^{3}P_{1}\rightarrow^{3}P_{0}$ & 609.8 & 2.3$\pm$0.1   & 3.1$\pm$0.1\tabularnewline

\hline

OH $^2\Pi_{3/2} 7/2^{-}\rightarrow7/2^{+}$ & 84.42$^a$& 9.1$\pm$0.7 & -- \\
OH $^2\Pi_{3/2} 7/2^{+}\rightarrow5/2^{-}$ & 84.60 &  7.4$\pm$0.7 & -- \\ 
OH $^2\Pi_{3/2} 5/2^{-}\rightarrow3/2^{+}$ &119.23 & 4.8$\pm$0.6 & -- \\
OH $^2\Pi_{3/2} 7/2^{+}\rightarrow5/2^{-}$ &119.44 & 5.8$\pm$0.6 & -- \\
OH $^2\Pi_{1/2} 3/2^{+}\rightarrow1/2^{-}$ &163.12 & 3.0$\pm$0.4 & --\\ 
OH $^2\Pi_{1/2} 3/2^{-}\rightarrow1/2^{+}$ &163.40 & 3.4$\pm$0.4 & -- \\

o-H$_2$O $2_{21}-1_{10}$&108.07    &6.5$\pm$0.9 &--\\
o-H$_2$O $4_{14}-3_{03}$&113.54$^b$&19.4$\pm$0.7 &--\\
o-H$_2$O $4_{23}-4_{14}$&132.41    &4.4$\pm$0.5 &--\\
o-H$_2$O $2_{11}-1_{01}$&179.53    &4.5$\pm$0.5 &--\\

CO34$\rightarrow$33 & 77.11 & 6.3$\pm$0.7 & --\\
CO31$\rightarrow$30 & 84.41$^a$ & 9.1$\pm$0.7 & --\\
CO30$\rightarrow$29 & 87.19 & 5.8$\pm$0.7 & --\\
CO29$\rightarrow$28 & 90.16 & 6.4$\pm$0.6 & --\\
CO28$\rightarrow$27 & 93.36 & 7.7$\pm$0.7 & --\\
CO24$\rightarrow$23 & 108.8 & 10.8$\pm$1.0\\
CO23$\rightarrow$22 & 113.5$^b$ & 19.4$\pm$0.7 & --\\
CO22$\rightarrow$21 & 118.6 & 14.5$\pm$0.6 & --\\
CO21$\rightarrow$20 & 124.2 & 14.1$\pm$0.4 & --\\
CO20$\rightarrow$19 & 130.4 & 15.7$\pm$0.5 & --\\
CO19$\rightarrow$18 & 137.2 & 12.8$\pm$0.5 & --\\
CO18$\rightarrow$17 & 144.8 & 16.1$\pm$0.6 & --\\
CO17$\rightarrow$16 & 153.3 & 19.5$\pm$0.6 & --\\
CO16$\rightarrow$15 & 162.8 & 16.0$\pm$0.4 & --\\
CO15$\rightarrow$14 & 173.6 & 19.9$\pm$0.6 & --\\
CO14$\rightarrow$13 & 186.0 & 16.6$\pm$0.9 & --\\
CO13$\rightarrow$12 & 200.5 & 14.9$\pm$1.7 & -- \\
CO12$\rightarrow$11 & 217.1 & 13.8$\pm$1.2 & -- \\
CO11$\rightarrow$10 & 236.8 & 13.8$\pm$0.8 & 2.2$\pm$0.7\\
CO10$\rightarrow$9  & 260.5 & 14.9$\pm$0.5 & 3.2$\pm$0.6\\
CO9$\rightarrow$8   & 289.4 & 13.9$\pm$0.3 & 2.5$\pm$0.4\\
CO8$\rightarrow$7   & 325.5 &  6.8$\pm$0.6 & 2.8$\pm$0.9\\
CO7$\rightarrow$6   & 372.0 &  7.5$\pm$0.3 & 3.6$\pm$0.4\\
CO6$\rightarrow$5   & 434.0 &  7.2$\pm$0.2 & 4.9$\pm$0.2\\
CO5$\rightarrow$4   & 520.7 &  9.0$\pm$0.1 & 5.1$\pm$0.1\\
CO4$\rightarrow$3   & 651.1 &  7.4$\pm$0.2 & 5.9$\pm$0.1\\

\hline                  
\end{tabular}
Notes: $^a$ blend of CO and OH; $^b$ blend of CO and H$_2$O.
\end{table}

The main transitions targeted with PACS and SPIRE are fine structure lines of [O\,I], [C\,I], [C\,II] and emission from CO, OH, and H$_2$O. The details of the detected transitions and their fluxes can be found in Table~\ref{t_emi_lines}. 

Overall, the R Mon spectra are richer in emission lines than PDS 27. Both R Mon and PDS 27 have a compact source in their surroundings, located at 64 and 58\arcsec, respectively. These 'neighbours' fall outside the PACS field of view, but they are in the field of view of SPIRE, as can be seen in Figure~\ref{f_spire_footprint}. The T Tauri companion of R Mon (at a distance of 0.69\arcsec) could in principle contribute to the spectra; however, we can neglect its contribution because of their small brightness ratio. Indeed, if we consider the [O\,I] 63 $\mu$m line flux and continuum of DG Tau, one of the brightest TTS, and scale it to the distance of R Mon, we obtain a line flux of 5.5$\times 10^{\rm{-17}}$ W/m$^2$ and a continuum flux of 0.45 Jy, thus 2\% and 0.5 \% of the line and continuum fluxes of R Mon, respectively, which is within the error of the observations.

\subsubsection{Oxygen fine structure lines}

Oxygen is the third most abundant element in the ISM and its fine structure line at 63.2 $\mu$m is by far the strongest line observed in R Mon, with a flux $\sim 242.1\pm0.5\times10^{-17}$ W/m$^2$. The [O\,I] line at 145 $\mu$m is also strong in R Mon and the ratio of [O\,I] 63 $\mu$m / [O\,I] 145 
$\mu$m is $9.9\pm0.2$. This is at the lower end of the 10-30 range found for Herbig Ae stars \citep{meeus2012A&A...544A..78M}, where the emission was attributed to the disc. It is also consistent with the median ratio of 10 (range from 5.5 to $>45$) found in deeply-embedded low-mass Class 0 protostars \citep{karska2013A&A...552A.141K}, where the emission is linked to the outflow. 

Assuming collisional excitation with H$_2$, \citet{2015ApJ...801..121N} find that the [O\,I] 63 $\mu$m / [O\,I] 145 $\mu$m ratio above 10 corresponds to densities in the 
$10^4-10^5$ cm$^{-3}$ regime, but is insensitive to the gas temperature. The values of the ratio around 9-10 match the models with $n\sim10^4$ cm$^{-3}$ when 
the main collisional partner is atomic H. The uncertainty in the H-to-H$_2$ fraction, however,
translates to an order of magnitude uncertainty in the derived densities.

On the other hand, the low values of the [O\,I] 63 $\mu$m / [O\,I] 145 $\mu$m ratio could be attributed to
line-of-sight absorptions in the [O\,I] 63 $\mu$m line. Recent velocity-resolved spectra of the [O\,I] 63 $\mu$m line show that self-absorptions are strong towards high-mass protostars \citep{2015A&A...584A..70L}, but negligible in their nearby, low-mass counterparts \citep{2017arXiv170407593K}. The spectral resolution of PACS ($\sim$ about 90 km/s at the 63 $\mu$m) is not sufficient to correct for potential self-absorptions. 

Finally, the considerably lower values found for the [O\,I] emission lines in PDS 27 suggest that the origin of emission may be different than in R Mon and low-mass protostars. We do not expect the [O\,I] 63 $\mu$m flux to be affected by off-position contamination, since the densities to excite the line are relatively high. We will discuss the possible scenarios for this source in Section \ref{ox+carbon}.

\subsubsection{Carbon fine structure lines}

In R Mon, the [C\,II] 157.7 $\mu$m line is clearly detected, and is present in all 25 spaxels, 
some of them even in absorption (see Figure~\ref{f_CII}). This is because the line is present 
with different strengths in the chop-on positions, as well as in the chop-off positions that are
subtracted to obtain the final spectra. In PDS 27, we see a faint absorption of [C\,II] 157.7 $\mu$m at the stellar position (see Figure~\ref{f_CII}). In the other spaxels we see stronger absorption features, especially towards the east. We can conclude that also in PDS 27 [C\,II] is present in variable amounts in the chop-on and -off positions. 

The [C\,I] (2-1) and (1-0) transitions at 370 and 610 $\mu$m are detected in both PDS 27 and 
R Mon. Both lines are seen in most of the beams of PDS 27, with exception of the noisiest
spectra. For R Mon, the [C\,I] (2-1) line is also omnipresent, while the lower-excitation
[C\,I] (1-0) line is only clearly present in the more central SLW beams (C3, C4, D2 and D3).

\subsubsection{Carbon monoxide}

Many transitions of carbon monoxide are detected in both PACS and SPIRE, as listed 
in Table~\ref{t_emi_lines}. In R Mon, PACS detects 16 transitions, from $J=34-33$ to $J=14-13$, 
with $E_{\text{up}}$ between 3279 and 581 K. On the other hand, no CO detections are seen with 
PACS for PDS 27. In R Mon, SPIRE detects 10 transitions, from $J=13-12$ to $J=4-3$, with $E_{\text{up}}$ between 503 and 55 K. For PDS 27, eight transitions are detected, from $J=11-10$ 
to $J=4-3$. 

For R Mon, the CO emission originates mainly from the central spaxel in PACS. In SPIRE, on 
the other hand, transitions up to CO\,(8-7) are observed in several beams. In particular, for 
the lowest transition, CO\,(4-3), we detect the line in most of the beams but the outer ones. 
The amount of beams in which a line is detected decreases for higher transitions, from roughly 
half (of 19) for CO\,(5-4) to only the five central ones covering the cavity (SLW C3, C4, C5, D3 
and D4) for CO\,(7-6). 

For PDS 27, the lowest transition CO\,(4-3) is also present in most of the beams, while for
increasing $J_{\rm{up}}$ the amount of beams decreases rapidly: only five for CO\,(5-4) and 
CO\,(6-5), three (SLW C3, C4 and C5) for CO\,(7-6) and two (SLW C3 and C4) at CO\,(8-7). 

For both objects, transitions higher than CO\,(8-7) are only observed at the position of the star.
Therefore, we can conclude that the CO emission, especially for transitions higher than (5-4) does
not originate from the larger-scale environment.

\subsubsection{OH and water}

HBe stars are more massive, warm and luminous than their HAe counterparts, and have stronger UV emission. \citet{thi2005A&A...438..557T} showed that the ratio H$_2$O/OH tends to be smaller when the ratio UV flux/gas density increases, probably due to the destruction of H$_2$O by UV radiation. In discs this means that H$_2$O could be located in deeper, colder layers than OH, but not be detectable, as \citet{fedele2011ApJ...732..106F} concluded. 

In R Mon we find four detections of water lines (E$_{\rm{up}}$ between 114 and 324 K), and three of the OH doublets (E$_{\rm{up}}$ between 120 and 291 K). There are also hints of doublets at 79.1, 71.2 and 65.2 $\mu$m, but they are not 3-sigma detections. These water lines detected are typically the strongest observed, as also seen in a sample of low-mass protostars \citep{karska2013A&A...552A.141K}. In this context it is important to note that in Herbig Ae stars, warm water lines were only convincingly seen in HD 163296 \citep[spectral type A1;][]{meeus2012A&A...544A..78M, fedele2013A&A...559A..77F}, while HD 100546 (spectral type B9.5Ve) only shows evidence of cold water \citep[observed with {\em Herschel}/HIFI;][]{hogerheijde2011Sci...334..338H}. It is surprising that no high-excitation (E$_{\rm{up}} >$ 500 K) water lines are detected, unlike what is observed in gas-rich T Tauri discs \citep{riviere2012A&A...538L...3R} and the Herbig Ae star HD 163296 \citep{meeus2012A&A...544A..78M}. For the three OH doublets detected in R Mon we constructed a rotational diagram which was fitted with a single temperature component, T$_{\text{rot}} \sim 190-340$ K.

In contrast, in PDS 27 we do not detect any H$_2$O or OH lines, likely because the UV radiation is too strong for molecules to survive photo-dissociation, at least in those layers where its emission would be visible to us.

\subsection{Dust features}

The forsterite feature at 69 $\mu$m, evidence for crystalline dust, has been observed 
in the PACS spectra of several HAeBe discs \citep[seven out of 23 observed;][]{sturm2013A&A...553A...5S}.
In R Mon and PDS 27 we do not see forsterite at 69 $\mu$m. This could indicate that forsterite 
is either absent, or at a too high/low temperature to emit efficiently at 69 $\mu$m, as is 
discussed by \citet{maaskant2015A&A...574A.140M}.

\section{Discussion}
\label{s_discussion}

\subsection{Physical conditions of the gas}

\begin{table}
\caption{Line flux ratios and highest CO transition detected. The star is located in C3, while 
the neighbour is in D4 for R Mon and C4 for PDS 27.}
\label{t_spire_beams}      
\centering   
\begin{tabular}{crrrc}       
\hline\hline                 
Beam & rCI   &  rCICO  &highest    \\
     &      &         &CO         \\
\hline
R Mon &       &        &           \\
C2   &$>$ 1.78           & --     &  --       \\
C3   &  1.4$\pm$0.1     &  0.007$\pm$0.001 &34-33    \\
C4   &  1.3$\pm$0.1    &   0.17$\pm$0.05 &8-7      \\
D4   &  2.0$\pm$0.3 &$>$ 0.28 &7-6      \\ 
\hline                       
PDS 27  &    &        &           \\
C2   &0.5$\pm$0.1  &$>$ 0.64 &4-3      \\
C3   &1.0$\pm$0.1  &   0.21$\pm$0.03 &11-10    \\
C4   &1.7$\pm$0.2  &   0.23$\pm$0.05 &8-7      \\ 
D3   &1.0$\pm$0.1  &$>$ 0.6 &5-4      \\
\hline
\end{tabular}
\end{table}

Line ratios of CO and CI lines are interesting since they are sensitive to the physical conditions of the emitting gas. In particular, they can be used to derive - at least comparatively -
the density and temperature of different regions 
\citep[e.g.,][]{kramer2004A&A...424..887K,rodon2015A&A...579A..10R}.
Both neutral and ionised carbon can be formed in low density gas, as they have critical densities of only $5--30\times10^{2}$ cm$^{-3}$ \citep{kaufman1999ApJ...527..795K}, while for the lowest CO transition in our data ($J=4-3$) the critical density is 100 times higher, $2.6\times10^{5}$ cm$^{-3}$.
 
We selected five representative SPIRE SLW beams to study the gas in the environment of our sources. 
For R Mon we selected C1, C2, C3 (R Mon), C4 and D4 (compact neighbour). For PDS 27 we selected C1, C2, C3 (PDS 27), C4 (neighbour) and D3. Apart from the central beam, the SPIRE spectra are not flux calibrated, but given that we want to calculate line ratios, this does not hamper our analysis. Also the fact that we do not have PACS spectra for the other beams does not affect our results, as we do not see higher$-J$ CO transitions in those beams. For both objects, no lines were detected in beam C1. For R Mon, the beam C2 has a low signal-to-noise spectrum, where only [C\,I] (2-1) was detected. We calculate the following line ratios, following \citet{rodon2015A&A...579A..10R}, see Table~\ref{t_spire_beams}:

\begin{description}
\item[rCI] = CI(2-1)/CI(1-0) 
\medskip
\item[rCICO] = $\Sigma$CI/$\Sigma$CO
\end{description}

The rCI ratio is sensitive to the density: optically thick [C\,I] lines have line ratios of about 1 or less, whereas optically thin lines have line ratios larger than 1. If the lines are optically thin, the excitation temperature $T_{\rm{ex}}$ is a function of the line ratio, with higher temperatures for higher ratios. For the beams at/around R Mon, we find rCI values between 1 and 2, indicating optically thin gas, while for PDS 27 we observe one beam (C2) where the gas is optically thick. Overall, we find higher CI line ratios in the parabolic shell above R Mon, suggesting warmer temperatures (as we describe in Section \ref{diagramas}) than around PDS 27.

The relative importance of cooling by CI and CO can be characterised by the ratio rCICO that takes all CI and CO lines into account (see Table~\ref{t_spire_beams}). A lower value of rCICO shows that cooling by CO is more important than by CI, a result of a higher density. In R Mon cooling by CO clearly dominates, less so in PDS 27. On the other hand, in R Mon beams C2 and D4, as well as in PDS 27 beams C2 and D4 cooling by CI dominates, thus in a lower-density environment. In the following subsection, 
we will use the CO rotational diagrams to further analyse the temperature of selected regions.

\subsection{CO rotational diagrams}
\label{diagramas}
\begin{figure}[t]
\begin{center}
\includegraphics[scale=0.21]{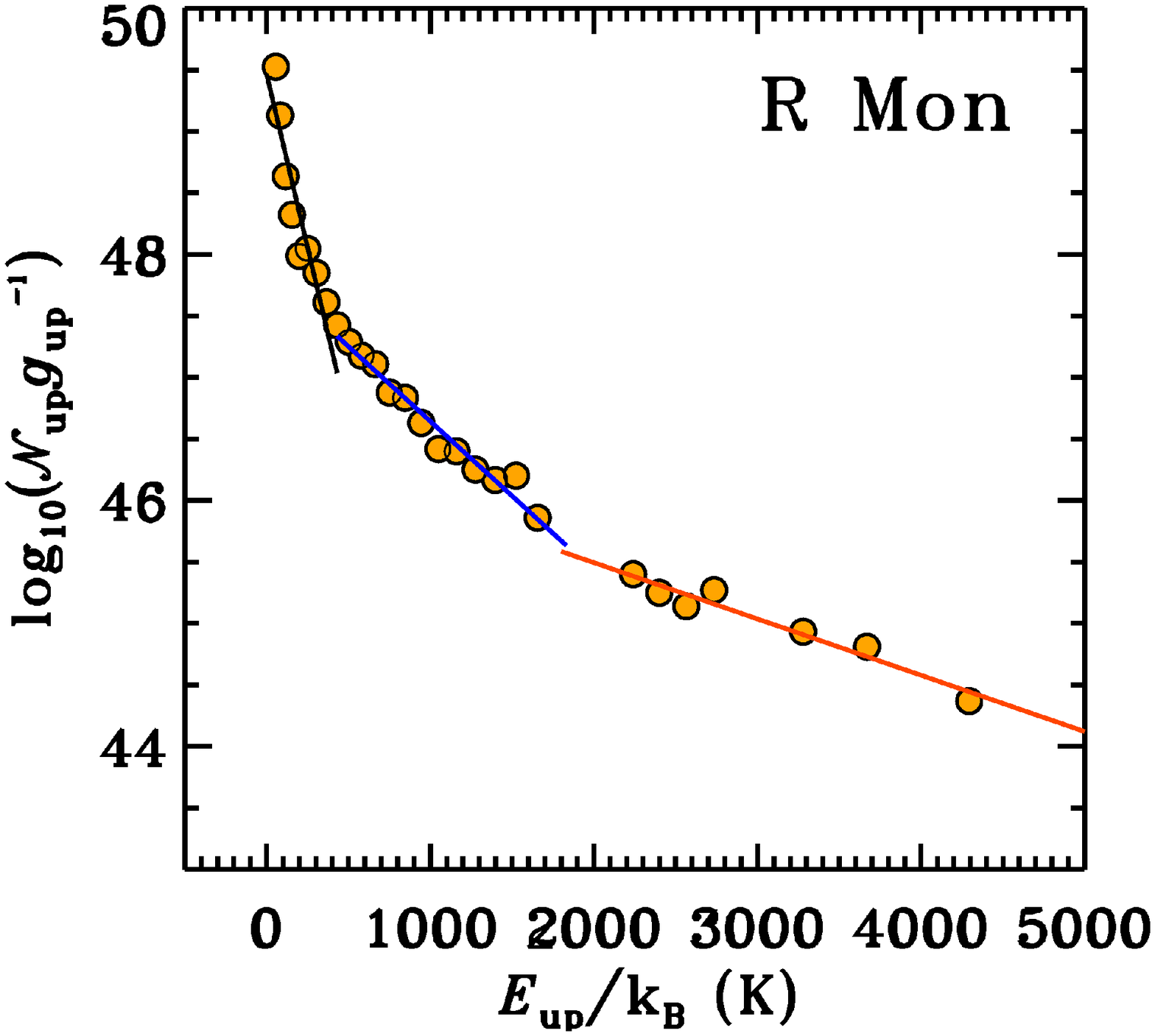}
\includegraphics[scale=0.21]{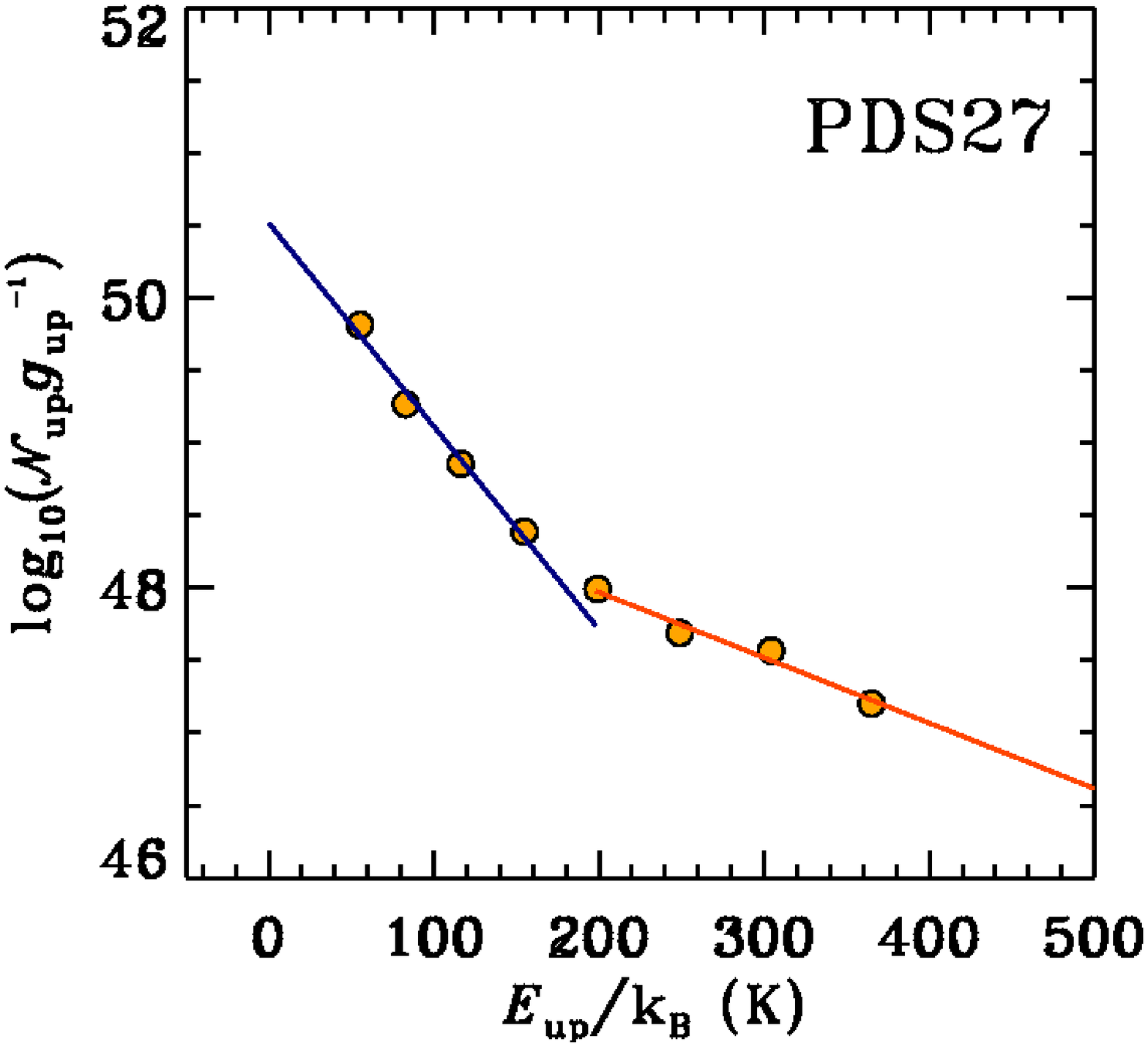}

\medskip

\includegraphics[scale=0.21]{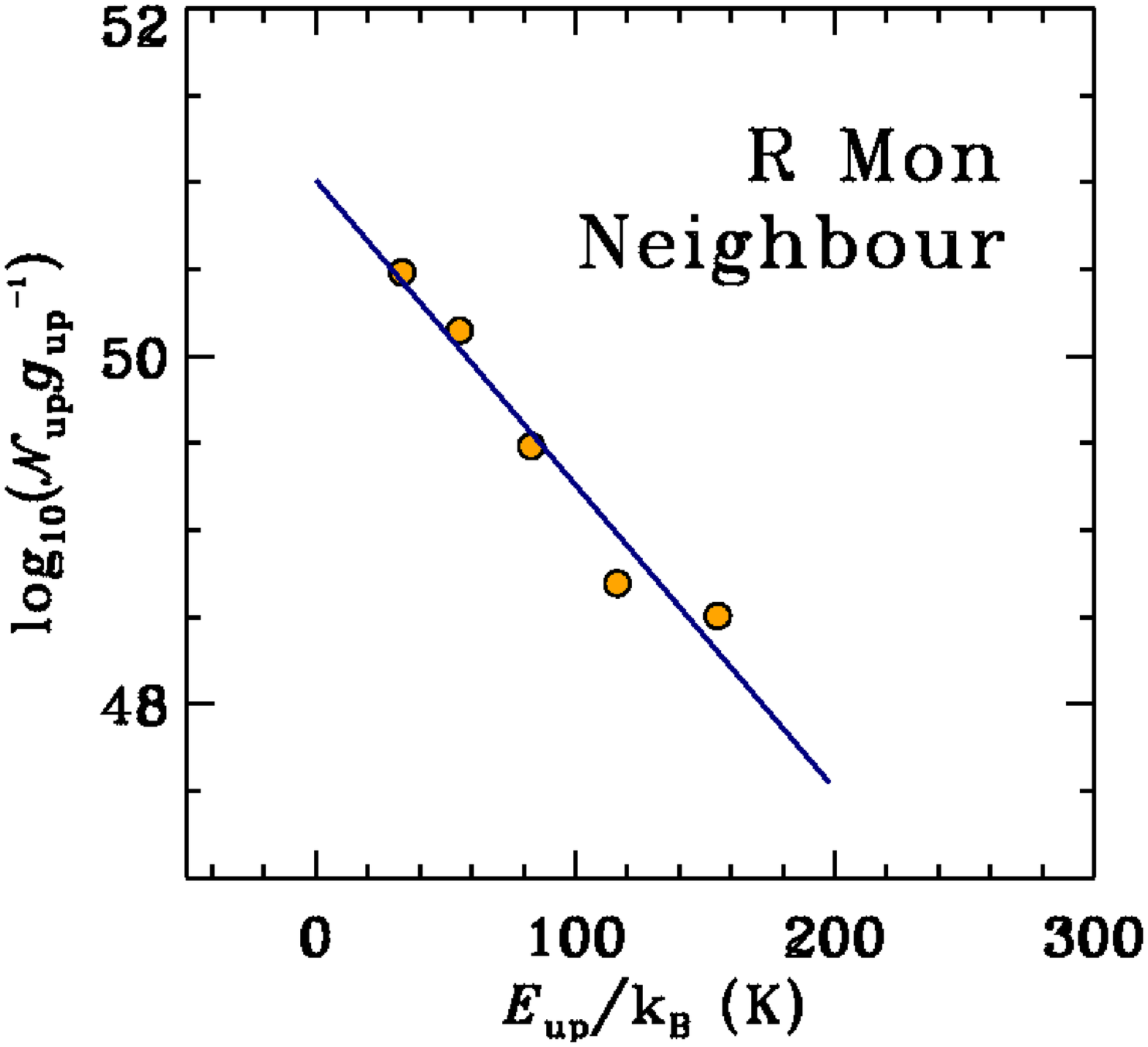}
\includegraphics[scale=0.21]{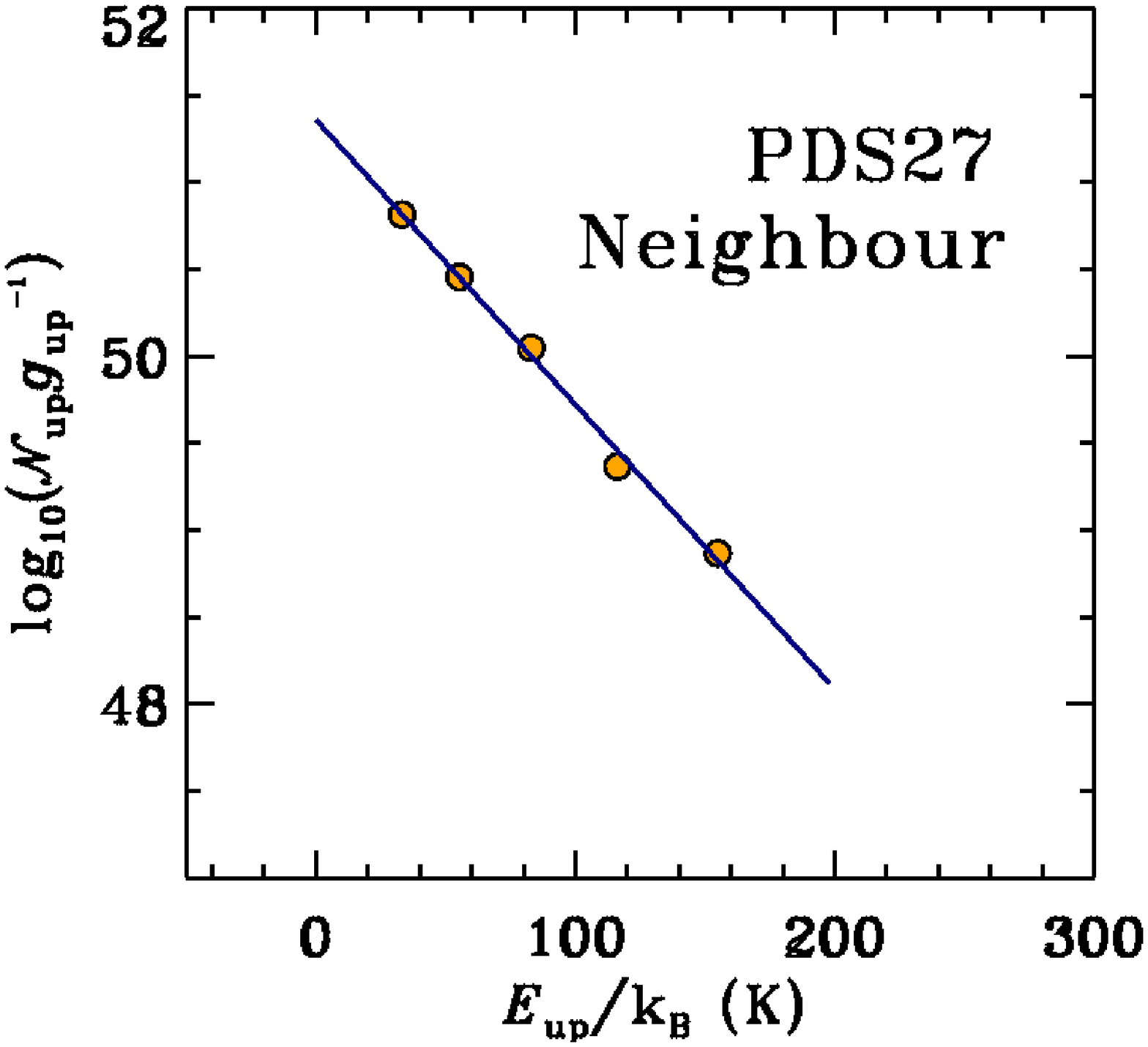}
\caption{CO rotational diagrams for R Mon, PDS 27 and their compact neighbours. The emission can be  divided into three different components for R Mon and two components for PDS 27: a cold component (for R Mon) is shown as a black line, whereas a warm and a hot component are represented by a blue and orange line, respectively. For the neighbours, only a single cold component is observed, represented by a blue line.}
\vspace{0.2in}
\label{f_rotdia}
\end{center}
\end{figure}

Carbon monoxide is the easiest molecule to analyse because its critical density, the upper level energy, and the frequency of its transitions increase monotonously as a function of the upper state, $J$. In our spectra, we detected 26 transitions in R Mon and eight for PDS 27. In Figure~\ref{f_rotdia}, we present the CO rotational diagrams for our sources and selected regions of their environment. Under the assumption of local thermodynamic equilibrium (LTE), and optically thin emission, the number of molecules per degenerate sub-level in the upper state of the transition, $N_\text{J}/g_\text{J}$, is proportional to the line flux. This way, one can assign a rotational temperature ($T_\text{rot}$) that would be expected to be equal to the kinetic temperatures if all levels were thermalised, so that the temperature of the emitting gas can be determined.

The rotational diagram of R Mon shows a positive curvature in the entire wavelength range. Such a 
shape can be the result of a smooth distribution of gas temperatures captured in the beam \citep{neufeld2012ApJ...749..125N,manoj2013ApJ...763...83M}. Alternatively, the CO diagram can be reflecting the presence of three distinct physical components.

\begin{table}
\caption{CO Rotational temperatures and total number of molecules found in the different 
components of the rotational diagrams. R Mon and PDS 27 are fitted with several components, the other selected regions (including the compact neighbours) can be fitted with a single temperature.} \label{t_trot}      
\centering          
\begin{tabular}{c c c c c}       
\hline\hline                
Source & $T_{\text{cold}}$ & $T_{\text{warm}}$ & $T_{\text{hot}}$ & $N_{\text{molecules}}$\tabularnewline
 & (K) & (K) & (K) &  \tabularnewline
\hline                        
R Mon & 77$\pm$12 & 358$\pm$20 & 949$\pm$90 & (9.1$\pm$2.8)$\times 10^{50}$\\
Neighbour & 25$\pm$8 & -- & -- & (1.0$\pm$0.4)$\times 10^{52}$\\
Beam C2 & 42$\pm$14 & -- & -- & (2.5$\pm$0.7)$\times 10^{50}$\\
Beam C4 & 32$\pm$10 & -- & -- & (6.1$\pm$0.4)$\times 10^{51}$\\
\hline
PDS 27 & 31$\pm$4 & 96$\pm$12 & -- & (4.1$\pm$1.2)$\times 10^{51}$\\
Neighbour & 27$\pm$6 & -- & -- & (2.3$\pm$0.3)$\times 10^{52}$\\
Beam C2 & 30$\pm$6 & -- & -- & (2.8$\pm$0.8)$\times 10^{51}$\\
Beam D3 & 29$\pm$5 & -- & -- & (4.2$\pm$0.9)$\times 10^{51}$\\
\hline                  
\end{tabular}
\end{table}

We fit three linear components to the CO rotational diagram of R Mon, with the breaks at $E_\text{up}$ = 431 K and $E_\text{up}$ = 1800 K. The temperatures corresponding to those components are $\sim$80 K, 360 K, and 950 K (see Table~\ref{t_trot}). The presence of warm and hot temperature components agrees with previous observations of HAeBe stars such as HD 100546 \citep[e.g.,][]{meeus2013A&A...559A..84M, vanderwiel2014MNRAS.444.3911V}. In PDS 27, we can fit two components, corresponding to rotational temperatures of $\sim$30 K and 100 K. For the neighbours, we only fit one component with a rotational temperature of 25 K (R Mon) and 27 K (PDS 27). The total number of CO molecules among the different components that is observed is 3.4$\times10^{50}$ for R Mon and 4.1$\times10^{51}$ molecules in PDS 27. Additionally we selected several beams in the surrounding environments of R Mon (SLW C2, D4) and PDS 27 (SLW C2, D3) with detections up to CO\,(8-7), and derived rotational temperatures in the range 30-40 K.

Rotational diagrams, however, do not provide a unique description of the gas conditions. Non-LTE molecular excitation models are necessary to reproduce the observed $T_{\rm{rot}}$ of the different components. R Mon is the only case where we find warmer components of 350-950 K, therefore we can compare these results with the NLTE models shown in Fig. 12 of \citet{karska2013A&A...552A.141K}. 
Based on the H$_2$O emission and the high densities needed to excite it, the range of possible scenarios is better described by a situation where CO is close to LTE in a warm, dense ($n(\rm{H}_2) \sim 10^6-10^9 \rm{cm}^{-3}$) gas. These high densities are compatible with highly compressed gas, which has been found at a few hundred AU of the protostar and at shock positions \citep[e.g.,][]{2012A&A...538A..45S, 2013A&A...551A.116T}.

In surveys of deeply-embedded low-mass protostars, CO emission can be described by up to 3 temperature components corresponding to 
$\sim$100, $\sim$300, and $\sim$700 K \citep{green2013ApJ...770..123G, karska2013A&A...552A.141K, 2013ApJ...763...83M}, and it is interpreted as arising in shocks \citep{2014A&A...572A...9K, 2017arXiv170407593K}. The  similarity of the CO emission in R Mon to those sources (Figure \ref{f_rotdia}) suggests that also here the bulk of emission originates from an outflow, and not the disc. On the other hand, the lack of detection of CO $J_{\text{up}}>10-9$ towards PDS 27 shows its similarity to the more evolved disc sources where both the CO and H$_2$O emission is scarce. This interpretation is further supported by the correlation between the [O\,I] 63 $\mu$m emission and continuum in 63 $\mu$m in disc sources 
(see Section \ref{ox+carbon}).

\subsection{[O\,I] and [C\,II]}
\label{ox+carbon}
\begin{figure}
\centering
\includegraphics[scale=0.37]{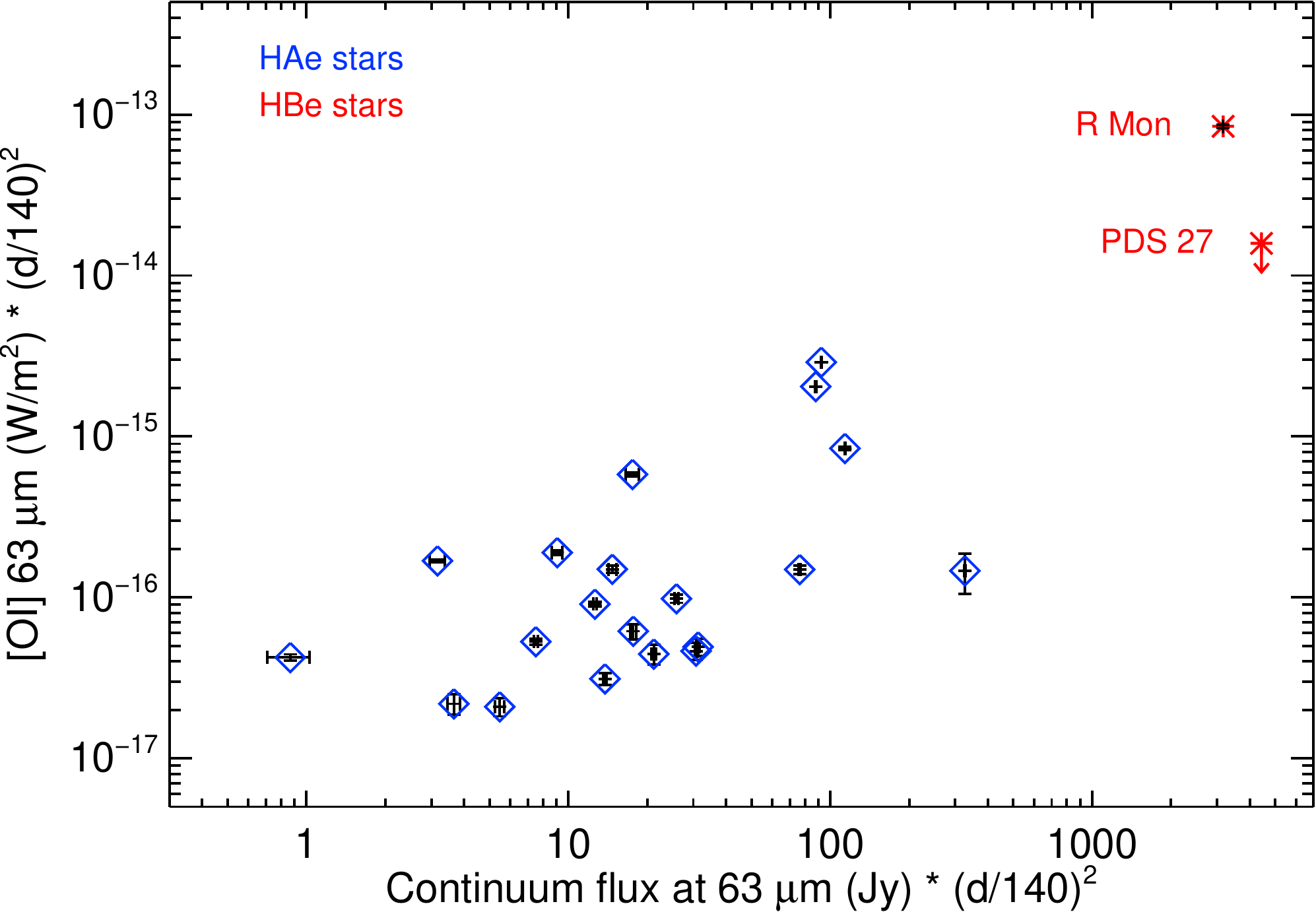}

\bigskip

\bigskip

\bigskip

\includegraphics[scale=0.37]{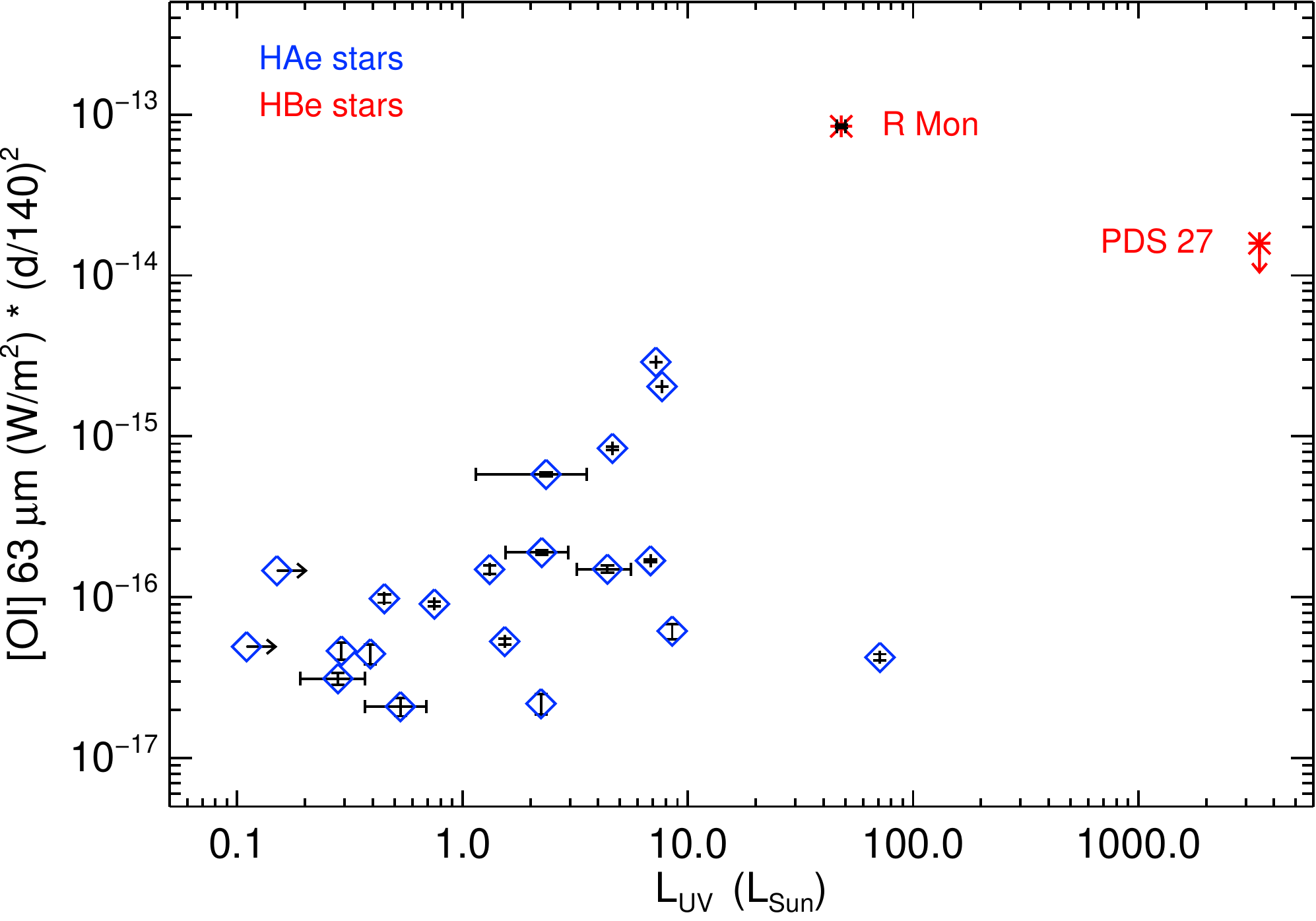}
\caption{[O\,I]63 $\mu$m versus continuum flux at 63 $\mu$m (top), and UV luminosity (115-243 nm, bottom). All the data were scaled to a common distance of 140 pc, to allow comparison. For PDS 27 we plot the upper limit.}
\label{f_OIcontUV}
\end{figure}

Previous studies have shown a correlation between the [O\,I] 63 $\mu$m line flux and the continuum flux at 70 $\mu$m for a sample of TTS \citep{howard2013ApJ...776...21H} and HAe stars \citep{meeus2012A&A...544A..78M}. \citet{howard2013ApJ...776...21H} also found that sources with an outflow have stronger [O\,I] line fluxes than their non-outflow counterparts (see their Fig. 16). We study the [O\,I] emission in our sources and compare it to the observed one in HAe stars \cite[data from GASPS observations;][]{meeus2012A&A...544A..78M} in Figure~\ref{f_OIcontUV}. The top panel shows the observed [O\,I] emission versus the continuum flux. We see that, while PDS 27 still could follow the trend of the HAe stars, the line flux of R Mon is too high, as has been observed in TTS with outflows. Most likely, part of its [O\,I] emission originates at the base of its outflow (given that the emission is only seen in the central spaxel; see Figure~\ref{f_CII}), and not just in the disc.

UV and X-ray radiation are very important for the chemistry and temperature balance of protoplanetary discs \citep[e.g.,][]{kamp2010A&A...510A..18K, aresu2012PhDT........81A}. 
In particular, UV photons can heat their discs through the photo-electric effect in PAHs in the surface layers, as well as photo-evaporate the outer disc \citep{gorti2009ApJ...690.1539G}. In Figure~\ref{f_OIcontUV}, bottom panel, we plot the [O\,I]\,63 $\mu$m line flux as a function of the UV luminosity (calculated as lower limits from atmosphere models in the range of 1150-2430 $\AA$). For HAe stars, there is a correlation between the [O\,I]\,63 $\mu$m flux and the UV luminosity, as reported earlier for a limited sample in \citet{pinte2010A&A...518L.126P} and supported by disc models in \citet{kamp2011A&A...532A..85K}. R Mon clearly lies above the expected [O\,I] flux, which can be a consequence of an increase in OH photo-dissociation in the disc surface and/or close to the star, therefore a PDR emission result from the UV irradiation of the outflow. In addition, the data presented in this paper -enhanced [O\,I], [C\,II] and CO lines, as well as the presence of water lines- show clear evidence for emission from the shocked material due to the presence of the outflow, which can add up to the [O\,I] emission from the disc. However separating the contribution of the outflow from the PDR emission remains hard without resolved emission profiles. As the stellar UV luminosity for PDS 27 is much higher, one could also expect a higher [O\,I] flux, but this is not the case. This is possibly due to the rapid dissipation of gas in the disc of PDS 27 due to its harsh UV radiation. 

The [C\,II] 157.7 $\mu$m emission line of ionised carbon can originate in the disc surface, where it is mostly optically thin \citep{kamp2011A&A...532A..85K}, but also in the surroundings of the disc. The interpretation of the [C\,II] 157.7 $\mu$m line is complex due to its relatively low spatial and spectral resolution. It could originate both in the disc and the environment, as the observed extended emission indicates. In PDS 27 we see the line in absorption, suggesting that the emission of the surroundings, used for background subtraction is higher than that of the star itself.

\subsection{Comparison with other sources}

\begin{table*}
\begin{center}

\caption{Total line luminosities in PACS and SPIRE lines (using only the observed transitions in Table \ref{t_emi_lines}), luminosity ratios and L$_{\rm{UV}}$. Line and continuum luminosities for HD 97048 and HD 100546 from \citet{meeus2012A&A...544A..78M,meeus2013A&A...559A..84M,fedele2013A&A...559A..77F,vanderwiel2014MNRAS.444.3911V}.}
\label{t_linelum}
\begin{tabular}{cccccccccc}
\hline \hline
Object &\multicolumn{6}{c}{Cooling Line luminosity (10$^{-3}$ L$_{\odot}$)}& L$_{\rm{mol}}$/L$_{\rm{[O\,I]}}$ & L$_{\rm{CO}}$/F$_{\rm{70}}$ & L$_{\rm{UV}}$ (L$_{\odot}$)\\
       & OI     & CI     & CII    & CO     & OH      &H$_2$O       &       &\\
\hline
R Mon  & 53.20  & 1.10   & 4.86   & 63.80  & 7.14    &6.94 &1.46   &3.3    &48\\
PDS 27 &$<$12.17& 3.00   &$<$2.53 & 14.71  & --      & --  &$>$1.21&0.54   &3450\\
\hline
HD 97048& 1.32  &$<$0.02 & 0.09   & 0.34   & 0.027   & --  &0.28   &6.4e-3 &7.69\\
HD 100546&1.94  &$<$0.01 & 0.06   & 0.47   & 0.026   & --  &0.26   &8.6e-4 &7.22\\
\hline
\end{tabular}
\end{center}
\end{table*}

The far-IR emission in protostars can originate in distinct parts of the stellar environment, as described by \citet{vandishoeck2011PASP..123..138V}.  In order to determine the dominant heating mechanism in our sources, we calculated the total line luminosity accounting for the observed transitions in each detected species (see Table~\ref{t_linelum}).

For R Mon, cooling by CO and [O\,I] dominates, with CO 20\% more than [O\,I]. Also, cooling by OH and H$_2$O is present in R Mon in equal amounts, while in PDS 27 only cooling by CO is found. We compare cooling in HBe with that of two early HAe stars, HD 100546 and HD 97048, which have the largest UV fluxes of the HAe sample in \citet{meeus2012A&A...544A..78M}. The HAe stars experience much less line cooling than the HBe stars, what can be explained by their difference in UV luminosity as shown in \citet{kamp2011A&A...532A..85K}. This is also shown by the ratio of molecular to atomic cooling, L$_{\rm{mol}}$/L$_{\rm{[O\,I]}}$, which is a factor 5 lower for HAe stars than for HBe stars: while in HAe stars atomic cooling dominates, HBe stars show more molecular cooling. Since our HBe stars are younger than the HAe stars and have a denser environment, it is interesting to compare to low-mass young objects: \citet{karska2013A&A...552A.141K} show that the ratio L$_{\rm{mol}}$/L$_{\rm{[O\,I]}}$ is close to unity for class I and II, while it is a factor 4-5 larger for class 0, the youngest sources, in agreement with our results. Our objects are somewhat more evolved, and in addition to the stellar heating of the disc, the UV-heated gas along cavity walls and winds can play a central role if shocks are present in HBe stars.

Outflow shocks can easily explain the heating of CO molecules, and therefore the unusually high CO luminosity in our HBe stars, the enhanced [O\,I] emission, and the presence of extended FIR water lines \citep[e.g.,][]{2010A&A...518L.120N, karska2013A&A...552A.141K} whose origin is corroborated by the broad line profiles of $\text{H}_2\text{O}$ from HIFI \citep[e.g.,][]{2012A&A...542A...8K, 2014A&A...572A..21M}. In particular, R Mon shows a powerful outflow of a few hundreds of km/s \citep{canto1981ApJ...244..102C, close1997ApJ...489..210C}. The CO line-to-continuum ratio is at least a factor of $\sim$1000 larger for HBe than for HAe stars, which suggests the presence of both UV-heated and shock-excited gas \citep{2013ApJ...762L..16M}. While the emission emission lines observed in R Mon can be explained with a combination of low-velocity shocks and PDRs \citep[e.g.][]{kamp2011A&A...532A..85K, 2013ApJ...779L..19P}, the environment in PDS 27 suggests that the central star dispersed most of the material and the detected emission comes mainly from the disc. On the contrary, no strong outflows are known in HAe stars, and the heating of the gas is mainly stellar.

\section{Conclusions}
\label{s_conclusions}

We presented and analysed {\em Herschel} PACS and SPIRE spectroscopy, covering the range $\sim $ 60 to 700 $\mu$m, to study the disc and environment of the Herbig Be stars R Mon and PDS 27. Our results can be summarised as follows: 

\begin{itemize}

\item We use new far-IR spectroscopy data to complement the ancillary near-IR and millimetre data for our targets in order to study their respective SEDs and calculate IR excesses. When compared with HAe stars, the IR excess of R Mon is much stronger, while that of PDS 27 is weaker, what could point to a smaller or less gas-rich disc. On the other hand, the FIR/NIR ratio of both stars is similar to that of group I HAe stars, indicative of a flared disc.

\medskip

\item 

[O\,I] 63 $\mu$m line is by far the strongest line observed in R Mon; with a [O\,I] 63/145 ratio of 9.9, at the lower end of the range found for HAe stars (10-30). In PDS 27 we do not detect [O\,I]. R Mon does not follow the correlation of [O\,I] 63 $\mu$m line with UV luminosity nor with 70$\mu$m continuum flux observed in HAe stars \citep{meeus2012A&A...544A..78M}.
[C\,II] is present in variable strengths in all the spaxels of the PACS detector. In R Mon it is seen in emission, and in PDS 27 in absorption, indicating that at the chop off position the flux is higher than at the star. 

\medskip

\item We detect four water lines and three OH doublets in R Mon, but none in PDS 27. We attribute the presence of water in R Mon to the shocked material from the outflow, and the absence of water and OH lines in PDS 27 to the harsh UV field, photo-dissociating those molecules.

\medskip

\item Forbidden lines of neutral carbon were detected for both sources at 370 and 609 $\mu$m. However, these lines are also detected in most other SPIRE beams - in a FOV of $\sim 3 \arcmin$, 
indicating that the lines originate in the low density environment rather than in the discs of the HBe stars.

\medskip

\item We detect many CO emission lines: 26 transitions in R Mon (with PACS+SPIRE), and eight in PDS 27 (SPIRE). For the neighbours we detect five transitions, each within the SPIRE range. The 
highest $J$ transition ($J=34-33$ with $E_{up}$ = 3279 K) is observed in R Mon. We construct rotational diagrams and find different temperature and density components in R Mon and 
PDS 27: a cold component with $T_{\rm{rot}}$ = 77 and 31 K for R Mon and PDS 27, respectively, and a warm component with $T_{\rm{rot}}$ = 358 and 96 K for R Mon and PDS 27, respectively. An extra hot temperature component of 949 K is observed for R Mon. For the neighbours we only need one component $\sim$ 25 K, thus cooler sources.

\medskip

\item The derived physical conditions and gas cooling budget in R Mon show a close similarity to more embedded star forming regions, suggesting that the emission originates from the surroundings rather than the disc. The dominant CO cooling, the enhanced [O\,I] and [C\,II] and the presence of water lines suggest that shocks associated to the outflow in the environment of R Mon, are the main source for the gas heating. In contrast, the FIR characteristics of PDS 27 resemble those of Herbig Ae stars. Here, the gas is UV-heated from the B2 star, dispersing the surrounding material; therefore the detected emission originates from the disc.

\end{itemize}

\begin{acknowledgements}
     The authors thank Daniel Harsono, Eric Pellegrini and Sacha Hony for excellent conversations that have improved this manuscript. GM is funded by Spanish grant RyC--2011--07920. GM and CE acknowledge partial support from the Spanish grant AYA2014--55840--P. AK acknowledges support from the Polish National Science Center grants 2013/11/N/ST9/00400 and 2016/21/D/ST9/01098.
\end{acknowledgements}


\bibliographystyle{aa}
\bibliography{HBe.bib}

\appendix

\onecolumn

\section{SPIRE beams} \label{app_spire}
\begin{figure*}
\centering
\includegraphics[scale=0.55]{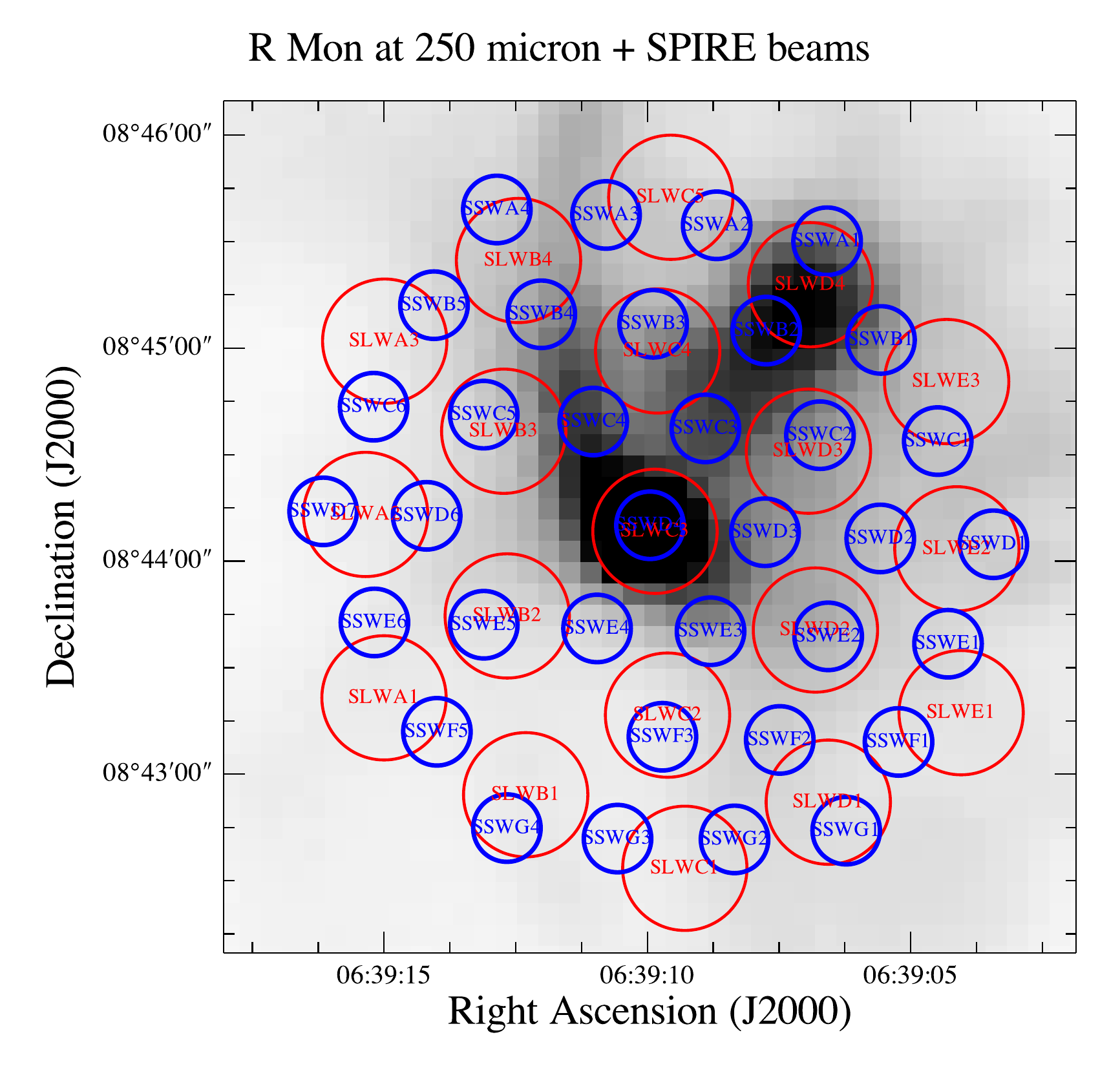}\\
\includegraphics[scale=0.55]{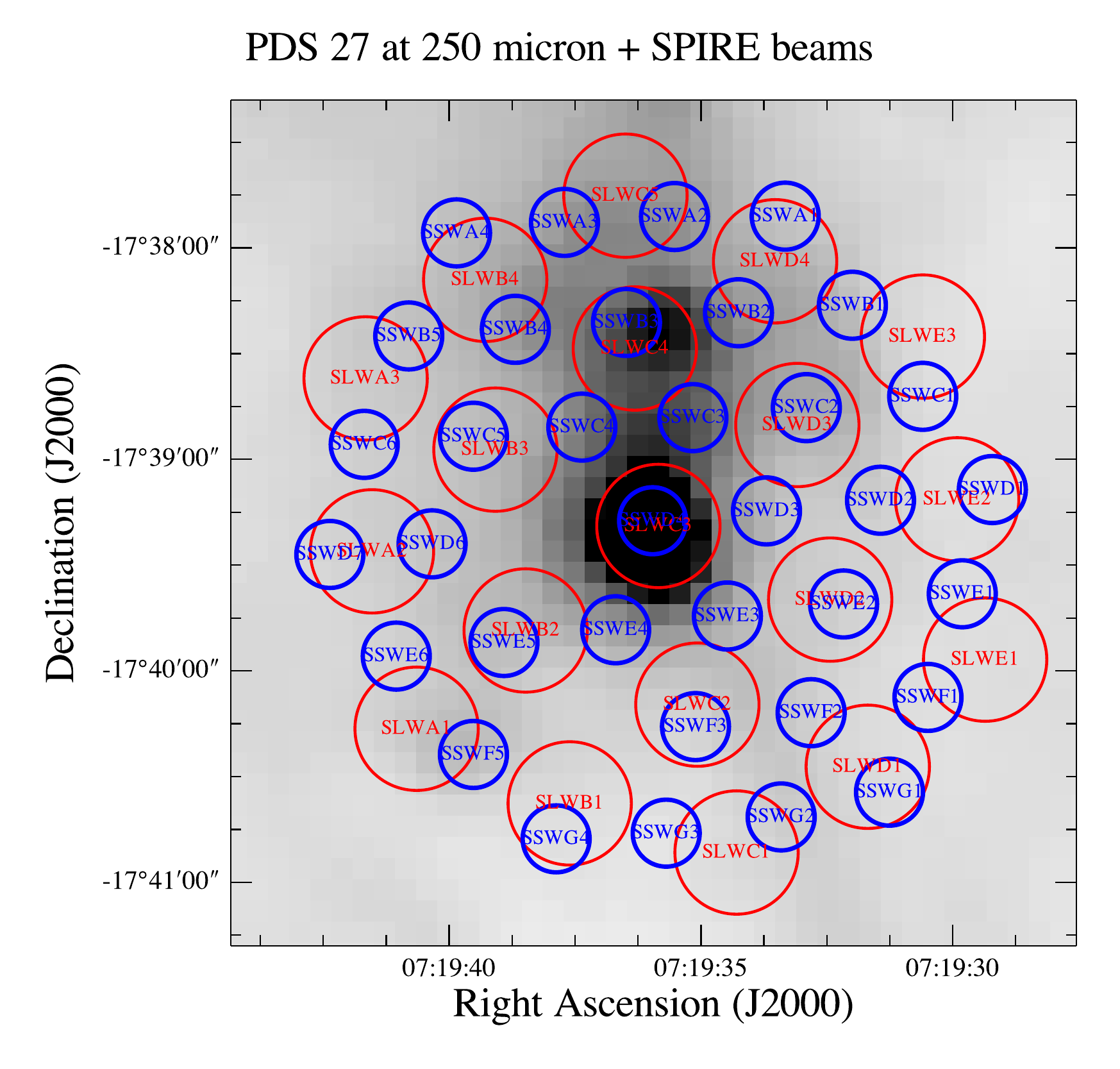}
\caption{SPIRE footprint. Shown are the beams for the SSW (red, large circles) and 
SLW (blue, small circles) detectors on top of the SPIRE image at 250 micron.}
\label{f_spire_footprint}
\end{figure*}

\begin{figure*}
\begin{center}
\includegraphics[scale=0.9]{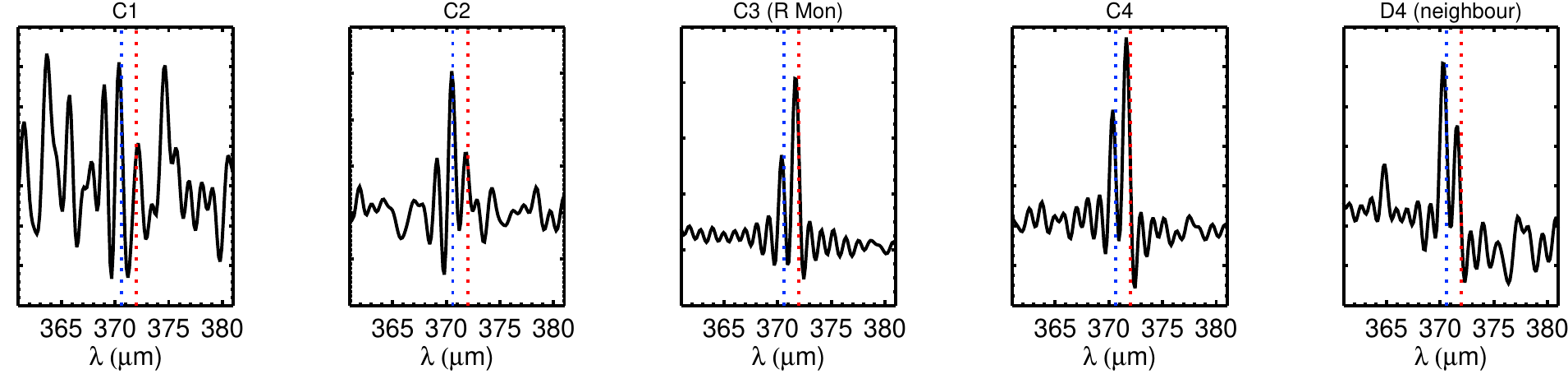}
\includegraphics[scale=0.9]{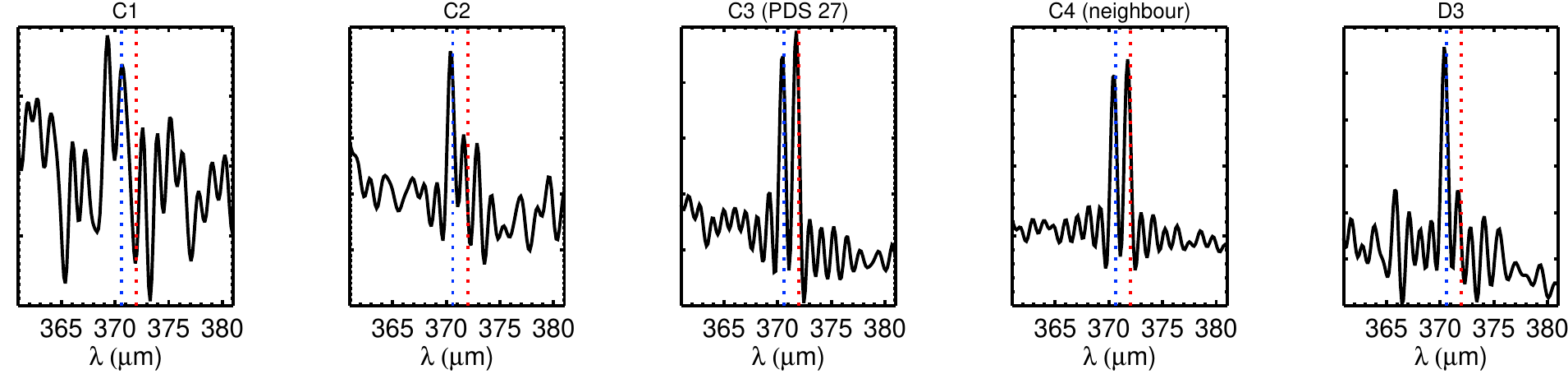}
\caption{The [C\,I] (2-1) and CO\,(7-6) lines at 370.6 and 372.0 $\mu$m line (blue and red) for selected beams around R Mon (top) and PDS 27 (bottom).}
\end{center}
\label{f_CICO}
\end{figure*}

In Figure~\ref{f_spire_footprint}, we plot the beams of the SPIRE detectors on the SPIRE
images, together with the corresponding labels. In Figure~\ref{f_CICO} we zoom in around 370 $\mu$m, to reveal what species 
dominates cooling at those wavelengths: in R Mon and beam C4, CO dominates, while in PDS 27 and C4 there are roughly equal contributions. In others beams
neutral carbon emission dominates.

\section{The [O\,I] and [C\,II] emission as observed with PACS} \label{app_pacs}
\begin{figure*}
\centering
\includegraphics[scale=0.25]{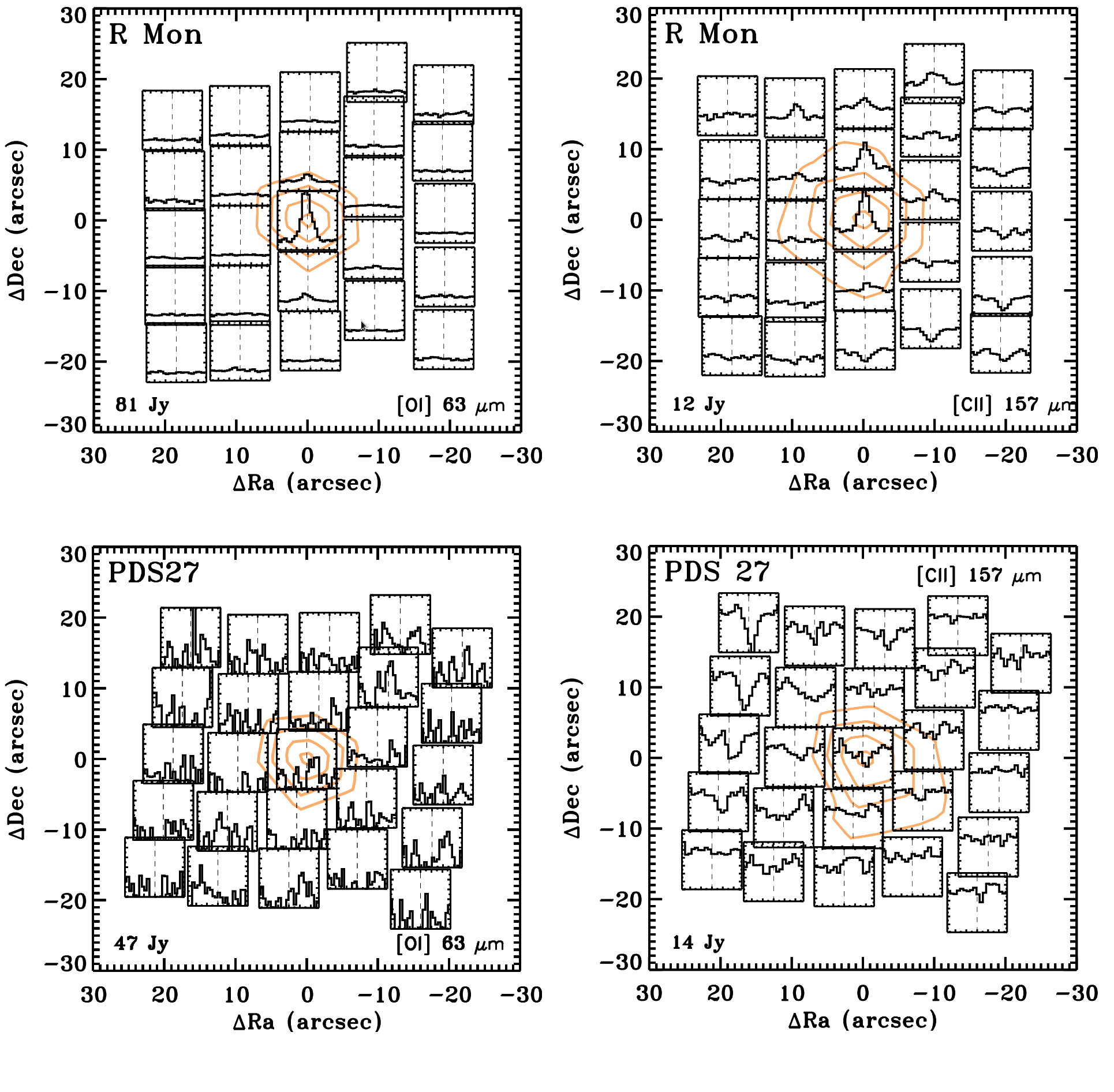}
\caption{The [O\,I]\,63 and [C\,II]\,157.7$\mu$m line for R Mon (top row) and PDS 27 (bottom row) }
\label{f_CII}
\end{figure*}

In Figure~\ref{f_CII} we show the 25 spaxels around the stars zoomed in
on the [C\,II] feature. It is clear that the feature is present in variable
strengths, in R Mon in emission at the stellar position and towards the 
N, while in absorption to the S. In PDS 27 we see the feature in absorption
in almost all spaxels.

\label{lastpage}
\end{document}